\documentclass[12pt,letterpaper,usenames,dvipsnames]{article}
\usepackage{jheppub}

\allowdisplaybreaks[2]  
\usepackage{pifont}
\usepackage{bm} 
\usepackage{mathrsfs}   
\usepackage{dsfont}  
\usepackage{ytableau}   
       
\usepackage{float}
\usepackage{hhline}
\usepackage{mathtools}
\usepackage{tabstackengine}
       \setstacktabbedgap{1ex}

\graphicspath{{figures/}}	 
\usepackage{tikz}
\usetikzlibrary{arrows,calc,snakes,shapes,shapes.arrows,
decorations.markings,decorations.pathmorphing}
\newcounter{sarrow}
\newcommand\xrsquigarrow[1]{\stepcounter{sarrow}%
\begin{tikzpicture}[decoration={snake,amplitude=.7mm},>=angle 90]
\node (\thesarrow) {\strut#1};
\draw[->,decorate] (\thesarrow.south west) -- (\thesarrow.south east);
\end{tikzpicture}}
\tikzset{>=triangle 45}
\tikzstyle{gr}=[draw,circle,green!50!black,fill=green!50!black,scale=.6]
\tikzstyle{Bl}=[draw,circle,blue,scale=.7]
\tikzstyle{R}=[draw,circle,fill=red,scale=.7]
\tikzstyle{bl}=[draw,circle,fill=black,scale=.2]
\tikzstyle{bbc}=[draw,circle,fill=black,scale=.75]
 
\usepackage{array} 
\usepackage{multirow} 
\usepackage{colortbl} 
\usepackage{arydshln} 
\usepackage{stfloats}  
\usepackage{cellspace}
\usepackage{xcolor}   

\usepackage{url} 
\usepackage{verbatim} 

\def\bar{\overline}
\def\til{\widetilde}

\def\del{{\partial}}

\def\vev#1{{\langle{#1}\rangle}} 
\newcommand{\beq}{\begin{equation}}
\newcommand{\eeq}{\end{equation}}
\newcommand{\bpm}{\begin{pmatrix}}
\newcommand{\epm}{\end{pmatrix}}
\newcommand{\bsm}{\begin{smallmatrix}}
\newcommand{\esm}{\end{smallmatrix}}

\def\^{\wedge}
\def\I{\mathds{1}}
\def\Tr{{\rm\, Tr}}

\def\rank{\text{rank}}

\def\Out{\text{Out}}
\def\U{{\rm\, U}}
\def\SU{{\rm\, SU}}
\def\suf{{\mathfrak{su}}}
\def\O{{\rm\, O}}
\def\SO{{\rm\, SO}}
\def\sof{{\mathfrak{so}}}
\def\SL{{\rm\, SL}}
\def\PSL{{\rm\, PSL}}
\def\GL{{\rm\, GL}} 
\def\Sp{{\rm\, Sp}}
\def\Spin{{\rm\, Spin}}

\def\C{\mathbb{C}} 

\def\I{\mathbb{I}}
\def\N{\mathbb{N}}

\def\R{\mathbb{R}} 
\def\Z{\mathbb{Z}} 

\def\gf{{\mathfrak g}}

\def\tu{{\til u}}

\def\bz{{\bf z}}

\def\cC{{\mathcal C}}

\def\cF{{\mathcal F}}
\def\cH{{\mathcal H}}
\def\cI{{\mathcal I}}

\def\cM{{\mathcal M}}
\def\cN{{\mathcal N}}
\def\cO{{\mathcal O}}

\def\cS{{\mathcal S}}
\def\cU{{\mathcal U}}
\def\cV{{\mathcal V}}
\def\cW{{\mathcal W}}

\def\a{{\alpha}}

\def\g{{\gamma}}
\def\G{{\Gamma}}
\def\tG{{\til\G}}
\def\d{{\delta}}
\def\D{{\Delta}}

\def\z{{\zeta}}
\def\th{{\theta}}

\def\l{{\lambda}}

\def\m{{\mu}}

\def\r{{\rho}}

\def\s{{\sigma}}

\def\S{{\Sigma}}
\def\t{{\tau}}

\def\f{{\phi}}
\def\vf{{\varphi}}

\def\w{{\omega}}
\def\-{{\text{-}}}

\title{Coulomb branches with complex singularities}
\author[1,2]{Philip C. Argyres,}
\author[3]{Mario Martone,}
\affiliation[1]{University of Cincinnati,
Physics Department, PO Box 210011, Cincinnati OH 45221}
\affiliation[2]{California Institute of Technology, Walter Burke Institute for Theoretical Physics, Pasadena CA 91125}
\affiliation[3]{University of Texas, Austin,
Physics Department, Austin TX 78712}
\emailAdd{philip.argyres@gmail.com}
\emailAdd{mariomartone@utexas.edu}

\abstract{
We construct 4d superconformal field theories (SCFTs) whose Coulomb branches have singular complex structures.  This implies, in particular, that their Coulomb branch coordinate rings are not freely generated.  Our construction also gives examples of distinct SCFTs which have identical moduli space (Coulomb, Higgs, and mixed branch) geometries.  These SCFTs thus provide an interesting arena in which to test the relationship between moduli space geometries and conformal field theory data.

We construct these SCFTs by gauging certain discrete global symmetries of $\mathcal N=4$ superYang-Mills (sYM) theories.   In the simplest cases, these discrete symmetries are outer automorphisms of the sYM gauge group, and so these theories have lagrangian descriptions as $\mathcal N=4$ sYM theories with disconnected gauge groups.
}
 
\begin{document}
\maketitle

\section{Introduction and summary}

The existence of moduli spaces of vacua with constrained complex structures for supersymmetric quantum field theories has provided a powerful tool for the exact computation of certain observables.  But the connection of moduli space complex geometry to the local operator algebra of the QFT is not obvious.  

For example, the relation between the conformal data of superconformal field theories (SCFTs) and their moduli space geometries is not yet systematically understood.    Scalar primary operators forming a chiral subring of the SCFT operator algebra are the natural candidates for the operators whose vevs parameterize the moduli space of vacua.  But, despite notable recent progress \cite{Beem:2017ooy,Hellerman:2017sur,Hellerman:2018xpi}, basic questions about this relationship are unanswered:  Is a necessary and sufficient condition for an SCFT to have a moduli space that it has a chiral subring?  Can the chiral ring have nilpotents?  Is the coordinate ring of the moduli space the reduced chiral ring?  (I.e., is the moduli space as a complex space given by the set of vevs of the chiral ring fields consistent with the ring relations?)  Is the special K\"ahler structure of Coulomb branches of the moduli spaces encoded in the local operator algebra of the SCFT, and if so, how?

As a step towards answering these questions, it is useful to find large classes of moduli space geometries which can be used to refine various conjectures about the relationship between conformal data and the complex goemetry of moduli space.  For instance, $\cS$-class \cite{Gaiotto:2009we,Gaiotto:2009hg,Chacaltana:2010ks}, geometric engineering \cite{Klemm:1996bj}, and F-theory \cite{Vafa:1996xn,gr1512} techniques permit the construction of large classes of Coulomb branch geometries of 4d $\cN=2$ SCFTs (among other things).  A regularity noted in \cite{Tachikawa:2013kta,Beem:2014zpa} is that in all these constructions the Coulomb branch is simply $\C^r$ as a complex space.  (We will call the complex dimension, $r$, the ``rank" of the Coulomb branch.)  Assuming the identification of the coordinate ring of the Coulomb branch with the chiral ring of Coulomb branch operators of the SCFT, this is equivalent to saying that the Coulomb branch chiral ring of the SCFT is freely-generated, i.e., is isomorphic to the polynomial ring $\C[z_1,\ldots,z_r]$.

We will construct a new class of 4d $\cN=2$ SCFTs with the property that their Coulomb branches have complex singularities, and so, in particular, their coordinate rings are not polynomial rings.  Our construction also gives examples of distinct SCFTs which have identical moduli space (Coulomb, Higgs, and mixed branch) geometries.  

It was noted in \cite{Argyres:2017tmj} that non-freely-generated Coulomb branch chiral rings allow the existence of Coulomb branch scaling dimensions less than one without violating the unitarity bound \cite{Mack:1975je} on scalar field dimensions in the SCFT.  It was further conjectured in \cite{Caorsi:2018zsq} that this is the \emph{only} case in which non-freely-generated CB chiral rings occur.  Our construction of CBs with complex singularities are all couter-examples to this conjecture.   

The new class of SCFTs we construct here thus provides an interesting arena in which to test the relationship between moduli space geometries and conformal field theory data.  This class is formed by gauging certain discrete global symmetries of known ``parent" SCFTs to form new ``daughter" SCFTs.  The essential ingredients of this construction were already discussed in \cite{gr1512,am1604}, mostly in the context of theories with rank-1 Coulomb branches.  Here we generalize it in a straight forward way to arbitrary rank.

The simplest family of theories in which to perform this construction are parent $\cN=4$ superYang-Mills (sYM) SCFTs with gauge Lie algebra $\gf$.  We focus on these examples in which the resulting daughter theories have $\cN=4$ or $\cN=3$ supersymmetry.  The extension to $\cN=2$ parent or daughter theories is discussed briefly and is straight forward in principle, but we leave it to future work.

A very simple case of the construction, which has a purely weakly-coupled description as a gauge theory, is the construction of daughter $\cN=4$ theories by gauging a discrete global symmetry, $\G$, of $\cN=4$ sYM with connected gauge group $G$ which act on the vector multiplets by outer automorphisms of $G$: $\G\subset\Out(G)$.  Thus these daughter theories are simply $\cN=4$ sYM theories with the disconnected gauge groups $G\rtimes\G$.

It is well known \cite{Seiberg:1997ax} that the moduli space of an $\cN=4$ sYM theory with connected gauge group with Lie algebra Lie$(G)=\gf$ is a flat orbifold of $\C^{3r}$ by an action of the Weyl group $\cW(\gf)$ and carries an $\cN=4$ version of a special K\"ahler structure reflecting the constraints of low energy centrally extended $\cN=4$ susy and EM duality.  It also carries information on the S-duality of the SCFT through the dependence of the special K\"ahler structure on the exactly marginal gauge coupling $\t$.  The $\cN=2$ Coulomb branch is a $\C^r/\cW(\gf)$ complex ``slice" of this moduli space.  Its holomorphic coordinate ring is the ring of polynomials in $r$ variables invariant under the action of $\cW(\gf)$, which turns out to be itself simply a ring of polynomials in $r$ variables.  That is, the invariants of $\cW(\gf)$ are generated by $r$ polynomials in the original variables without further relations.  
The Coulomb branch of the daughter theory is then the orbifold $\C^r/[\cW(\gf)\rtimes\G]$ where the $\G$ action on $\C^r$ is worked out in this paper.  The holomorphic coordinate rings of these orbifolds are also described below, and are shown in many cases to not be freely-generated.

Other cases of this construction are where the discrete symmetry $\G$ does not commute with the whole $\cN=4$ algebra, but only an $\cN=3$ or $\cN=2$ subalgebra.  In this case the daughter theory is a strongly-coupled SCFT with no exactly marginal local operators.  We discuss the simplest of these cases, namely the ones preserving an $\cN=3$ superconformal symmetry, finding similar results for the complex structure of their Coulomb branches as in the $\cN=4$ cases.

The rest of the paper is organized as follows.  Section \ref{symms} is devoted to the construction of discrete global symmetries of $\cN=4$ sYM theories which commute with at least an $\cN=3$ supersymmetry.  Section \ref{gauging} then describes the orbifold structure of the moduli space of the resulting daughter theories upon gauging the discrete symmetries of the previous section.  Section \ref{Cstructure} reviews some useful objects, namely the Molien series of an orbifold coordinate ring and its plethystic logarithm, which can be computed algorithmically.  This enables one to obtain explicit information about the generators and relations of the Coulomb branch coordinate ring, and, at least in many low-rank examples, to determine the ring completely.  Section \ref{examples} then uses this machinery to compute in examples, illustrating cases of distinct SCFTs sharing identical moduli spaces, of Coulomb branches with complex singularities which are complete intersections, and ones with singularities which are not complete intersections.  Finally, section \ref{questions} concludes with comments on the generalization of our construction to theories with only $\cN=2$ supersymmetry, as well as a list of some open questions.

\bigskip

\emph{Note added:}  When this paper was being completed, \cite{Bourget:2018ond} appeared which substantially overlaps with our work.  In particular, that paper also describes $\cN=4$ sYM theories with disconnected gauge groups given by extensions of connected groups by outer automorphisms and further extends it to $\cN=2$ gauge theories as well.  Where our results overlap, they agree. We also learned from E. Pomoni, T. Bourton and A. Pini of an upcoming work \cite{Elli} with overlaps with our work. In particular in \cite{Elli} the index of many of the theories analyzed here is computed. Again we find agreement with our results when they overlap. We thank the authors for sharing the draft in advance.

\section{$\cN{=}3$-preserving discrete symmetries with CB action}\label{symms}

Our goal is to construct new ``daughter" $\cN=3$ SCFTs with different Coulomb branch (CB) geometries by gauging discrete symmetries of ``parent" SCFTs.  These symmetries must therefore preserve $\cN=3$ supersymmetry and act non-trivially on the CB of the parent theories.  The only continuous global symmetry which acts on the CB is the $\U(3)_R$ symmetry, so by definition a discrete subgroup of the $\U(3)_R$ does not leave the  $\cN=3$ supercharges invariant.  So a discrete symmetry that will do the job does not obviously exist.

Nevertheless, if the parent theory has enhanced $\cN=4$ supersymmetry, there do exist discrete symmetries, $\G$, which commute with an $\cN=3$ supersymmetry but which have a nontrivial action on the CB.  This was pointed out in the case of a free $\cN=4$ $\U(1)$ gauge theory by Garc\'ia-Etxebarria and Regalado in \cite{gr1512} as part of their string S-fold realization of $\cN=3$ SCFTs.  Their observation was generalized to $\cN=4$ $\SU(2)$ gauge theory and, more conjecturally, to (non-lagrangian) rank-1 $\cN=3$ and also further to $\cN=2$ theories by the authors in \cite{am1604}.  

We will review the identification of these symmetries and generalize them to parent $\cN=4$ theories with arbitrary rank $r>1$ CBs.  The result, which is similar to the rank-1 case described in \cite{am1604}, is that an $\cN=4$ sYM theory with simply-laced gauge Lie algebra $\gf$ has at most four such symmetries:
\begin{align}\label{Gs}
\begin{array}{llcc}
\G\  \simeq\ & \Z_k\quad & \qquad \t_* \qquad\ & \text{SUSY preserved}\\[1mm]
\hline \\[-3mm]
\G_2 & \Z_2 & \text{any}   & \cN=4 \\[2mm]
\G_3 & \Z_3 & e^{i\pi/3} & \cN=3 \\[2mm]
\G_4 & \Z_4 & i                  & \cN=3 \\[2mm]
\G_6 & \Z_6 & e^{i\pi/3} & \cN=3 
\end{array}
\end{align}
On the left are the names we give these symmetries; they are all $\Z_k$, $k=2,3,4,6$, groups.\footnote{More precisely, these are their subgroups which act faithfully on bosonic fields and EM charges of states on the moduli space of vacua.  Sometimes $\Z_{2k}$ is the group acting faithfully on the full set of fields.} $\t_*$ denotes the value of the gauge coupling of the $\cN=4$ sYM theory for which this symmetry occurs.  The last column shows the amount of supersymmetry these symmetries commute with.

The story is a bit more complicated for non-simply-laced $\gf$.  The classification \eqref{Gs} turns out also to work for $\gf=\sof(2r+1)$ and $\mathfrak{sp}(2r)$ but is not correct for the exceptional non-simply-laced Lie algebras $\gf=G_2$ or $F_4$.  What happens in these cases will be indicated below in footnotes.

Not all the symmetries in \eqref{Gs} are necessarily present for every $\cN=4$ sYM theory.  Such a theory with a given simple gauge Lie algebra, $\gf$, is specified by some further discrete data, namely the global form of the compact gauge Lie group \cite{Goddard:1976qe}, as well as by a choice of the spectrum of line operators \cite{Aharony:2013hda}.   These discrete choices affect whether and which of the $\G_k$ with $k>2$ are symmetries, as will be explained below.  

The $\cN{=}4$-preserving $\G_2$ is the outer automorphism group of the gauge group for all $\cN=4$ sYM theories.  This symmetry is non-trivial only for $\suf(N)$, $\sof(2N)$, and $E_6$ gauge Lie algebras.
It coincides with charge conjugation symmetry for the $\suf(N)$, $\sof(4N+2)$, and $E_6$ gauge  algebras, but is something different for $\sof(4N)$ algebras.

\subsection{Constructing the symmetries}\label{construct}

A key observation of \cite{gr1512} is that at special values of the gauge coupling, certain discrete subgroups, $\S_R\subset SL(2,\Z)$, of the S-duality group of an $\cN=4$ sYM theory are global symmetries which act non-trivially on the supercharges.  Thus, at these couplings, some S-duality identifications supply ``extra" discrete R-symmetries.

Following the discussion in \cite{gr1512,am1604}, we look for a   symmetry, $\G$, preserving at least an $\cN=3$ supersymmetry and acting non-trivially on the CB of the $\cN=4$ sYM theory generated by an element  
\begin{align}\label{GERsymm}
C := (\r,\s) \quad \in \quad \SU(4)_R \times \S_R .
\end{align}
Here $\SU(4)_R$ is the continuous R-symmetry group of the $\cN=4$ sYM theory.  Since we are looking at finite $\G$ generated by a single element, we will have $\G \simeq \Z_k$ for some $k$.  These will turn out to be the only possibilites.  

Since $\G$ is finite, $\r$ must be of finite order and so is a semisimple element of $\SU(4)_R$.  Then up to conjugation in $\SU(4)_R$, $\r$ can be chosen to be in a maximal torus.  Using the equivalence $\SU(4) \simeq \Spin(6)$, $\r$ can be represented by a simultaneous rotation in three orthogonal planes in $\R^6\simeq \C^3$:
\begin{align}\label{N4rho}
\r = \bpm e^{i\psi_1} &&\\ &e^{i\psi_2}&\\ &&e^{i\psi_3} \epm
\in \U(3) \subset \SU(4)_R .
\end{align}
The six real adjoint scalar fields, $\f^I$, $I\in{\bf 6}$ of $\SU(4)_R$, of the $\cN=4$ vector multiplet can be organized into a triplet of complex scalars, $\vf^a$, $a\in{\bf 3}$ of $\U(3)$, by defining $\vf^a = \f^{2a-1}+ i \f^{2a}$.  Then $\r$ acts as
\begin{align}\label{ractvf}
\r:\ \vf^a \to e^{i\psi_a} \vf^a .
\end{align}
The four chiral supercharges, $Q_\a^i$, $i\in{\bf 4}$ of $\SU(4)_R$, transform under $\r$ by the phases
\begin{align}\label{rhoQact}
\r:\ \begin{cases}
Q_\a^1 &\to e^{i(+\psi_1+\psi_2+\psi_3)/2}\ Q_\a^1 \\
Q_\a^2 &\to e^{i(+\psi_1-\psi_2-\psi_3)/2}\ Q_\a^2 \\
Q_\a^3 &\to e^{i(-\psi_1+\psi_2-\psi_3)/2}\ Q_\a^3 \\
Q_\a^4 &\to e^{i(-\psi_1-\psi_2+\psi_3)/2}\ Q_\a^4
\end{cases}\ .
\end{align}

An $\cN=4$ sYM theory with simple gauge Lie algebra, $\gf$, has an exactly marginal coupling, $\t$, taking values in the complex upper half-plane, and identified under S-duality transformations which form a finite-index subgroup $\cS\subset\SL(2,\Z)$.\footnote{\label{nslg}This is only true for simply-laced $\gf$. In the non-simply laced case $\SL(2,\Z)$ is replaced by the infinite discrete subgroup $H_q\subset\SL(2,\R)$ generated by $T = \left(\bsm 1&1\\0&1\esm\right)$ and $S_q = \left(\bsm 0&-1/q\\ q&0\esm\right)$ where $q$ is the ratio of the lengths of long to short roots of $\gf$ \cite{Dorey:1996hx,Argyres:2006qr}.}  In particular, under the action of an element, $\s$, of the S-duality group the sYM coupling transforms as 
\begin{align}\label{SLtrt}
\s:\ \t\to\frac{a\t+b}{c\t+d},
\qquad \text{if} \quad \s :=
\bpm a&b\\ c&d\epm \in \SL(2,\Z) .
\end{align}
S-duality transformations also transform the chiral supercharges by a phase \cite{Kapustin:2006pk}
\begin{align}\label{SLtr}
\s:\  Q_\a^i \to e^{i\chi} Q_\a^i
\qquad \text{where} \qquad
e^{i\chi} = \left(\frac{|c\t+d|}{c\t +d}\right)^{1/2} .
\end{align}
$\chi$ is only defined up to shifts by $\pi/2$ since such a shift is in the center of the $\SU(4)_R$ symmetry.  It is convenient to specify $\chi$ unambiguously by choosing that shift so that $-\pi/2 \le \chi < 0$.  Finally, the S-duality transformations have trivial actions on the vector multiplet scalars for simply-laced $\gf$.\footnote{\label{G2F4CBact}They have a non-trivial action described in \cite{Argyres:2006qr} when $\gf=G_2$ or $F_4$.} 

$\s \in \SL(2,\Z)$ can only be a symmetry of a theory at values of its coupling $\t$ fixed by the action of $\s$.   Suppose $\t_*$ is the value of $\t$ fixed by the action \eqref{SLtrt}.  Simple algebra then shows that $(c\t_*+d)$ satisfies the characteristic equation for $\s$, and is thus an eigenvalue of $\s$.

For $\S_R$ to be a discrete symmetry group, it must be a finite subgroup of $\SL(2,\Z)$, and so any $\s\in\S_R$ must have finite order.  Thus $\s$ must be diagonalizable and have eigenvalues which are conjugate roots of unity.  This can only happen if the discriminant of its characteristic polynomial is non-positive, which implies its trace (being an integer) takes one of the five values $\Tr\s\in\{-2,-1,0,1,2\}$, corresponding to elements of orders $\{2,3,4,6,1\}$, respectively.  Their conjugacy classes in $\SL(2,\Z)$ are 
\begin{align}\label{SigRs}
\begin{array}{llc}
\S_R\quad\ &[\s_k]\in[\SL(2,\Z)] & [\t_*] \\[1mm]
\hline \\[-3mm]
\Z_2 & \s_2 = -I & \text{any} \\[2mm]
\Z_3 & [\s_3] = [-ST]\ \text{or}\ [(-ST)^{-1}]\quad \  & e^{i\pi /3} \\[2mm]
\Z_4 & [\s_4] = [S]\ \text{or}\ [S^{-1}] & i \\[2mm]
\Z_6 & [\s_6] = [ST]\ \text{or}\ [(ST)^{-1}] & e^{i\pi /3}
\end{array}
\end{align}
where square backets denote conjugacy classes, and where $S=(\bsm 0&-1\\1&0\esm)$ and $T=(\bsm 1&1\\0&1\esm)$ generate $\SL(2,\Z)$.  The order of $\s_k$ is thus $k$, and the third column describes the $\SL(2,\Z)$ orbit of the value of the coupling fixed by $\s_k$ by giving its value in a fundamental domain of the $\SL(2,\Z)$ action on the upper half-plane.  Since $\s_3$ and $\s_6$ fix a different $\t_*$ than $\s_4$, and since the groups the $\s_k$ generate are related by $\Z_2 \subset \Z_4$ and $\Z_2\times\Z_3=\Z_6$, they cannot be combined to form other finite subgroups of $\SL(2,\Z)$ fixing a common $\t_*$.  Thus \eqref{SigRs} lists all the possible types of discrete subgroups of the S-duality group that can occur as symmetry groups.\footnote{\label{nslSduality}This conclusion is modified for non-simply-laced $\gf$.   Since $S_{\sqrt2}$ interchanges $\gf=\sof(2r+1)$ and $\mathfrak{sp}(2r)$ it is an equivalence between different theories, so there is no value of the coupling where it is a symmetry.  It follows that the maximum subgroup of $H_{\sqrt2}$ which can contain symmetries is the (Hecke) congruence subgroup $\G_0(2)\subset\SL(2,\Z)$ whose finite-order subgroups are $\Z_k$ for $k=2, 4$, related by $\Z_2\subset\Z_4$.  When $\gf = G_2$, the finite order subgroups of $H_{\sqrt3}$ are $\Z_k$ for $k=2, 4, 6, 12$, related by $\Z_2\subset \Z_4$ and $\Z_2 \ltimes \Z_6 = \Z_{12}$.  When $\gf = F_4$, the finite order subgroups of $H_{\sqrt2}$ are $\Z_k$ for $k=2,4,4',8$, related by $\Z_2 \subset \Z_4$ and $\Z_2 \ltimes \Z_4' = \Z_8$.  Finally, as we will show below, all the $\Z_2$ $\S_R$'s act trivially in these theories, so can be discarded.}  

Note, however, that not all of the possibilities in \eqref{SigRs} may occur for a given $\cN=4$ sYM theory.  The reason is that the S-duality group $\cS$ is not necessarily all of $\SL(2,\Z)$ but may be some finite-index subgroup, which might not contain elements of all these orders.  Furthermore, a given $\cS$ might also have multiple distinct copies of a given $\Z_k$ with each copy fixing a different value of $\t_*$.  (These different $\t_*$'s will all be in the same $\SL(2,\Z)$ orbit, as indicated in \eqref{SigRs}, but will be in distinct orbits of $\cS\subset\SL(2,\Z)$.)  The $\Z_2$ center of $\SL(2,\Z)$ appearing in \eqref{SigRs} is always part of the S-duality group but in some cases may be part of the gauge group (as we will explain below), and so act trivially.  Finally, note that if $\Z_3$ exists as a subgroup of the S-duality group fixing some $\t_*$, then there is also a $\Z_6=\Z_3\times\Z_2$ fixing it.  

It then follows from \eqref{SLtr} and the observation that $c\t_*+d$ is an eigenvalue of $\s_k$ that, irrespective of the specific S-duality group $\cS\subset\SL(2,\Z)$ that a theory realizes, if $\cS$ contains an element of order $k$, then it acts on the supercharges as\footnote{This remains true in the non-simply-laced cases, but now the possible values of $k$ are $k=4$ for $\sof(2N+1)$ or $\mathfrak{sp}(2N)$; $k=4,6,12$ for $G_2$; and $k=4,8$ for $F_4$.}
\begin{align}\label{SLtr1}
\s_k:\  Q_\a^i \to {\rm e}^{-i\pi/k}Q_\a^i .
\end{align}
This is slightly inaccurate: the $k$ appearing in $\s_k$ on the left of \eqref{SLtr1} is not necessarily the same $k$ appearing in the phase on the right, although they are always drawn from the same setof possibilities.  Depending on the eigenvalue of $\s_3$ realized by $c\t_*+d$, either the $k=3$ or $k=6$ phase may appear on the right in \eqref{SLtr1}; the same is true of $\s_6$.  But, as noted above, in any theory either both or neither of $\s_3$ and $\s_6=-\s_3$ occur as symmetries, and if one contributes a $k=3$ phase in \eqref{SLtr1}, the other contributes the $k=6$ phase.  Thus the set of phases realized in the possible $\S_R$ symmetry actions on the supercharges given by the rule \eqref{SLtr1} is correct even if the labelling of the generator as $\s_k$ is incorrect.  Since all we will use in the sequel is the action on the supercharges, we will henceforth label them using \eqref{SLtr1}, and can safely ignore the fact that the corrrespondence to S-duality elements given in \eqref{SigRs} might be permuted.

Now we want to find a $\r:=\r_k$ for each $\s_k$ in \eqref{SLtr1} such that the combined action of the pair $C_k:=(\r_k,\s_k)$ preserves at least an $\cN=3$ supersymmetry.  We start by constructing such symmetries which commute with the full $\cN=4$ supersymmetry.  

\subsection{$\cN{=}4$-preserving symmetries}\label{N4toN4symm}

Up to the action of the Weyl group of $\SU(4)_R$ (which permutes the $\psi_a$ and shifts any pair of them by $\pi$), it is not hard to see from \eqref{rhoQact} and \eqref{SLtr1} that the only way for the combined action of $(\r_k,\s_k)$ to leave all four supercharges invariant is to choose $k=2$ and 
\begin{align}\label{N4rhosols}
\r_2 := \{ \psi_1 = \psi_2 = \psi_3 = \pi \} 
\end{align}
in the representation \eqref{N4rho}, i.e., $\r_2 = -I\in \U(3) \subset \SU(4)_R$.  Then $\G_2 \simeq \Z_2$ generated by $C_2:= (\r_2,\s_2)$ is a discrete global symmetry of an $\cN=4$ sYM theory at all values of the coupling since $\s_2$ does not fix $\t$ \eqref{SigRs}.

In this case, since $\G_2$ is a symmetry even at weak coupling, it can be identified directly as a symmetry of the $\cN=4$ sYM theory lagrangian.  Since it is generated by a transformation which changes the sign of the electric and magnetic charges of states on the moduli space, it must change the sign of the Cartan subalgebra components of the vector field.  To be a $\Z_2$ symmetry of the  sYM action, it must extend to an involutive automorphism of the whole gauge Lie algebra.  The automorphism, $\s_2$, that does this is called the ``Chevalley involution" of $\gf$ \cite{Fuchs}, is unique up to conjugation by an inner automorphism, and extends to an involution of any Lie group $G$ with Lie$(G)=\gf$.  In a Chevalley-Serre basis of $\gf$ given by $\{ H^i, E^i_\pm,\ i=1,\cdots,\rank(\gf)\}$ write the gauge field components as
\begin{align}\label{}
A^\m = Z^\m_i H^i + W^{\m\pm}_i E^i_\pm + \ldots,
\end{align}
so the $Z^\m_i$ are the $\U(1)^{\rank(\gf)}$ gauge fields on the moduli space, the $W^{\m\pm}_i$ are the $W$-bosons of the $\suf(2)$ subalgebras associated to simple roots, and the remaining terms are the $W$-bosons associated to the other roots whose generators are constructed from commutators of the $E^i_\pm$.  Then the Chevalley involution action on $A^\m$ is determined by the action
\begin{align}\label{Chevinvol}
\s_2:\ Z^\m_i \mapsto - Z^\m_i, \qquad 
W_i^{\m\pm} \mapsto - W_i^{\m\mp} ,
\end{align}
on the simple $\suf(2)$ subalgebras, and extends uniquely to all components of $A^\m$ to respect the Lie algebra bracket and linearity.  

The involution defined by \eqref{Chevinvol} is not unique, but can be composed with any inner automorphism of $\gf$ to give another Chevalley involution.  But any inner automorphism is just conjugation by a gauge group element, which is a space-time independent gauge transformation of $A^\m$, so this family of Chevalley involutions are gauge equivalent to one another.  Note also that any choice of Cartan subalgebra of $\gf$ can be mapped to any other by such a gauge transformation.  So in every choice of Cartan subalgebra, $\s_2$ is gauge equivalent to $Z_i \mapsto -Z_i$, and the Chevalley involution is the unique involution with this property.

Recalling that the action of inner automorphisms on a given choice of Cartan subalgebra of $\gf$ defines the Weyl group, $\cW(\gf)$, it follows that $\s_2$ is an outer automorphism of $\gf$ if and only if $-I \notin \cW(\gf)$.  This is the case and, it is easy to check, only the case if $\gf$ has complex representations.   Thus $\s$ is an outer automorphism only for
\begin{align}\label{ChevOuter}
\s_2\in\Out(\gf) \quad \text{for}\quad
\gf = \suf(N)\ (N\ge 3), \quad
\sof(4N+2)\ (N\ge0), \quad 
\text{or}\quad E_6.
\end{align}
The $\sof(2)$ case of this list is just the free Maxwell theory originally discussed in \cite{gr1512}.

The above discussion identifies $\s_2$ as the action of a charge conjugation symmetry on Yang-Mills fields.  (While the fact that charge conjugation acts as the Chevalley involution on Yang-Mills fields surely must be known to experts, we could not find it described in standard field theory texts; hence the above discussion.)  The fact that it acts trivially (i.e., is a gauge transformation) for all simple Lie algebras except those listed in \eqref{ChevOuter} implies that the $\Z_2$ center of the $\SL(2,\Z)$ duality group of $\cN=4$ sYM theories acts trivially for gauge algebras not listed in \eqref{ChevOuter} --- their S-duality groups must thus be a subgroup of $\PSL(2,\Z)$ instead.

For $\G_2$ to commute with $\cN=4$ supersymmetry, it must act in this way on the whole  $\cN=4$ vector multiplet.  Thus
\begin{align}\label{Gam2}
C_2: \ 
(A^\m, \psi_\a^{i}, \vf^{a})_A \mapsto 
(\s_2)_A^B \cdot (A^\m, \psi_\a^i, \vf^a)_B,
\end{align}
where $A,B$ are Lie algebra indices and $(\s_2)_A^B$ is the map determined by \eqref{Chevinvol}.  From \eqref{Chevinvol} it clearly acts as 
\begin{align}\label{Gam2CBact}
C_2: \ \vf^a_i \to -\vf^a_i, \qquad i=1,\ldots,\rank(\gf)
\end{align}
on the Cartan subalgebra and thus on the moduli space.

Of the simple Lie algebras not in the list \eqref{ChevOuter}, only $\gf = \sof(4N)$ have outer automorphisms.  These give discrete symmetries preserving $\cN=4$ supersymmetry just as in \eqref{Gam2} but with $(\s_2)^B_A$ replaced by any representative of the outer automorphism action on the Lie algebra.\footnote{We thank J. Distler for emphasizing this point to us.}  These symmetries are not constructed from a generator of the form \eqref{GERsymm}.  Their existence suggests that symmetries of the form \eqref{GERsymm} constructed above and listed in \eqref{Gs} may not exhaust the list of all possible $\cN{=}3$-preserving discrete symmetries acting on the CB at strong coupling.  Some strategies for searching for such possible additional symmetries will be discussed in section \ref{questions}.

In the sequel we will consider the effects of gauging the outer automorphism symmetries of $\sof(4N)$ $\cN=4$ sYM on their CB geometries.  To that end, we will need an explicit action of the outer automorphism on a Cartan subalgebra and thus on the moduli space.   The outer automorphism group of $\sof(4N)$ is $\Z_2$ which can be thought of as acting as the symmetry of its Dynkin diagram, from which it follows that an action on the Cartan subalgebra can be taken to be\footnote{Here it is convenient to use a basis of the Cartan subalgebra in which the Killing metric is diagonal, proportional to $\d^{ij}$, instead of to the Cartan matrix.  In this basis the Weyl group is generated by permutations on the $i$ index and by sign flips of an even number of the $\vf_i$ fields.}  
\begin{align}\label{tGam2CBact}
\til C_2: \ 
\begin{cases}
\vf^a_i \to +\vf^a_i, & i=1,\ldots,2N-1 \\
\vf^a_i \to -\vf^a_i, & i=2N 
\end{cases}
\qquad \text{for}\qquad
\gf = \sof(4N) .
\end{align}
We will denote also by $\G_2$ this $\Z_2$ symmetry of the $\sof(4N)$ sYM theory generated by $\til C_2$.

In the special case of $\gf=\sof(8)$ there are additional outer automorphisms forming the permutation group on three elements, $S_3 = \Z_3 \rtimes \Z_2$.  In a simple basis of the Cartan subalgebra, a generator of the $\Z_2$ subgroup can be taken as in \eqref{tGam2CBact} (for $N=2$) while a generator of the $\Z_3$ subgroup is
\begin{align}\label{tGam3CBact}
\til C_3:\ 
\bpm \vf^a_1 \\ \vf^a_2 \\ \vf^a_3 \\ \vf^a_4 \epm
\mapsto
\frac12 \bpm +1 & +1 & +1 & -1 \\[.7mm]  
+1 & +1 & -1 & +1 \\[.7mm]
+1 & -1 & +1 & +1 \\[.7mm]
+1 & -1 & -1 & -1 
\epm
\bpm \vf^a_1 \\ \vf^a_2 \\ \vf^a_3 \\ \vf^a_4 \epm 
\qquad\text{for}\qquad \gf=\sof(8).
\end{align}
So for the $\sof(8)$ theory there are three possible inequivalent $\cN{=}4$-preserving discrete symmetries acting on the CB: the $\G_2 \simeq \Z_2$ generated by $\til C_2$, a $\tG_3 \simeq \Z_3$ generated by $\til C_3$, and a non-abelian $\tG_6 \simeq S_3$ generated by both $\til C_2$ and $\til C_3$.

\subsection{$\cN{=}3$-preserving symmetries}\label{N4toN3symm}

Up to the action of the Weyl group of $\SU(4)_R$, there is just one inequivalent choice of $\r_k$ for each $\s_k$ for $k=3,4,6$ in \eqref{SigRs} which preserves three supersymmetries, given by
\begin{align}\label{N3rhosols}
\r_k &:= \{ \psi_1 = \psi_2 = {+2\pi/k}, \ 
\psi_3 = {-2\pi/k} \} , \qquad k\in\{3,4,6\}
\end{align}
in the representation \eqref{N4rho}.  The combined $C_k := (\r_k,\s_k)$ action on the supercharges preserves an $\cN=3$ supersymmetry by leaving $Q^i_\a$ for $i=1,2,3$ invariant.  In this case the $C_k$ action on the vector multiplet scalars in a Cartan subalgebra of $\gf$ is\footnote{Except for $\gf=G_2$ or $F_4$; see footnote \ref{G2F4CBact}.} from \eqref{N4rho}
\begin{align}\label{GamkCBact}
C_k: \ 
\begin{cases}
\vf^a_i \to \exp\{ +2\pi i/k \} \vf^a_i
& a=1,2 \\
\vf^a_i \to \exp\{ -2\pi i/k \} \vf^a_i
&a=3
\end{cases}
\qquad
i=1,\ldots,\rank(\gf) .
\end{align}

Then $\G_k \simeq \Z_k$ generated by $C_k$ for $k=3,4,6$ are possible discrete global symmetries of an $\cN=4$ sYM theory at the fixed values of the coupling determined by \eqref{SigRs}.  Since these symmetries only occur at strong coupling, they are not apparent as symmetries of the sYM lagrangian, as $\G_2$ was.  Nevertheless, knowledge of the S-duality groups of $\cN=4$ theories allows us to determine when these symmetries exist (and act non-trivially).  Generally, the S-duality group is some finite-index subgroup of $\SL(2,\Z)$.  This subgroup can be determined as in \cite{Goddard:1976qe,Aharony:2013hda} by keeping track of the action of $\SL(2,\Z)$ generators on not just the gauge coupling, but also the discrete data specifying the sYM theory.  That data is the gauge Lie algebra, $\gf$, the choice of global form of the gauge group, $G$, and a maximal set of mutually local line operators.  There is a unique simply-connected compact Lie group $\til G$ with Lie$(\til G)=\gf$.  All other compact $G$ with the same Lie algebra are given by $G_i = \til G/\Pi_i$ for $\Pi_i \subset Z(\til G)$ a subgroup of the (finite, abelian) center, $Z(\til G)$ of $\til G$.  
For a given choice of $G_i$ there are roughly $|\Pi_i|$ choices of line operator spectrum \cite{Aharony:2013hda}.

For example, when $N$ is square-free, i.e., a product $N=\prod_{i\in I} p_i$ of distinct primes $p_i$, the possible global forms of the gauge group for $\gf=\suf(N)$ are $ \SU(N)/\Z_M$ where $M \mid N$.  As shown in \cite{Aharony:2013hda}, all these groups and their associated spectra of line operators are permuted by the S-duality group which is the congruence subgroup $\G_0(N) \subset \SL(2,\Z)$.   The number of elements of order 2 and 3 in $\G_0(N)$ considered as a subgroup of $\PSL(2,\Z)$ are known \cite{ModularForm1989}, from which it follows that there are elements of $\G_0(N)$ in $\SL(2,\Z)$ of order
\begin{align}\label{}
&k=4 &&\text{iff} && \forall i\in I \quad 
p_i=1\ (\text{mod }4) \quad \text{or}\quad p_i=2,\nonumber\\
&k=3\text{ and }k=6 &&\text{iff} && \forall i\in I \quad 
p_i=1\ (\text{mod }3) \quad \text{or}\quad p_i=3.\nonumber
\end{align}
Thus, as examples, among the first twenty-four square-free $N$'s, $\cN{=}4$ sYM with $\gf = \suf(N)$ for $N=6,11,14,15,18,22,23,30,33,35$ have no $\Z_{k>2}$ symmetries, for $N=2,5,10,17,26,29,34$ have only a $\Z_4$ symmetry, for $N=3,7,19,21,31$ have only $\Z_{3,6}$ symmetries, and for $N=13,37$ have all the $\Z_{3,4,6}$ symmetries.

As another set of examples, when $\gf = \suf(N^2)$ then there is a gauge group and choice of spectrum of line operators, denoted by $[\SU(N^2)/\Z_N]_0$ in \cite{Aharony:2013hda}, which has the full $\SL(2,\Z)$ group as its S-duality group.  These theories therefore all have $\Z_3$, $\Z_4$, and $\Z_6$ symmetries. 

\section{Gauging the symmetries}\label{gauging}

We now gauge these discrete symmetries of $\cN=4$ sYM theories.  This will project out all local operators of the theory which are not invariant under the symmetry.  Thus if some of the supercharges are charged under the symmetry, gauging the symmetry will reduce the amount of supersymmetry.  Also, the OPE algebra of local operators of the SCFT will be similarly projected.  But, since there are no dynamical gauge bosons associated to this gauging, the counting of multilocal operators remains essentially the same, and so the SCFT OPE coefficients like the $a$ and $c$ central charges which effectively count the local degrees of freedom (or enter into the OPE of energy momentum tensors) remain the same under discrete gauging.

The geometry of the moduli space of vacua of the theory will change under gauging if any of the fields getting vevs on the moduli space are charged under the discrete symmetry.  We will discuss in this subsection precisely how the moduli space geometry changes.  We start by reviewing the moduli space of vacua of $\cN=4$ sYM theories.

\subsection{Geometry of $\cN=4$ sYM moduli space}

The moduli space of vacua of $\cN=4$ sYM theories are parameterized by the vevs of the complex Cartan subalgebra scalar fields, $\vf^a_i$ for $a=1,2,3$ and $i=1,\ldots,r=\rank(\gf)$.  The geometry gets no quantum corrections so is locally flat $\C^{3r}$, but is orbifolded by any gauge identifications of a given Cartan subalgebra of the gauge Lie algebra.  These identifications are given by the finite Weyl group, $\cW(\gf)$, of the Lie algebra.  $\cW(\gf)$ acts as a real crystallographic reflection group on the real Cartan subalgebra, i.e., via orthogonal transformations, $w\in \O(r,\R)$, with respect to the Killing metric on the Cartan subalgebra.  Thinking of the vector multiplet scalar vevs, $\vf^a_i$, as linear coordinates on $\C^3 \otimes_\R \R^r$, the Weyl group acts as $I_3 \otimes w$ matrices, where $I_3$ denotes the $3\times 3$ identity matrix.  With this action, an $\cN=4$ sYM with gauge Lie algebra $\gf$ has the moduli space
\begin{align}\label{N4sYM}
\cM(\gf) = \C^{3\,r} /\cW(\gf) ,
\qquad r:= \rank(\gf).
\end{align}
Note that this result does not depend on the other discrete data (global form of the gauge group, spectrum of line operators) defining the sYM theory.

Geometrically $\cM(\gf)$ is a flat orbifold.  More precisely, in a basis of the Cartan subalgebra where the Killing form is the Cartan matrix, $C^{ij}$, of $\gf$, then the hermitean metric is $h = C^{ij} \d_{ab} d\vf^a_i d\bar\vf^b_j$ locally, with orbifold singularities at the fixed loci of the $\cW(\gf)$ action (which occur in real codimension 6).  In this basis, the $I_3\otimes w$ linear action of the Weyl group on the $\vf^a_i$ coordinates is represented by an integral matrix, $w \in \GL(r,\Z)$, reflecting the crystallographic property of the Weyl group.

Since the massless degrees of freedom on $\cM(\gf)$ are the $\U(1)^r$ Cartan subalgebra gauge fields, it is a Coulomb branch.  In particular, it carries an $\cN{=}4$ analog of a special K\"ahler structure in which the complex $\vf^a_i$ are the (analog of) special coordinates, and $\vf^{ai}_D := i C^{ij} \vf^a_j$ are dual special coordinates.  $\vf_D$ and $\vf$ transform in the $2r$-dimensional representation of the low energy EM duality group,
\begin{align}\label{N4EMmonod}
\bpm \vf^a_D \\ \vf^a \epm
\ \xrsquigarrow{\small{$\ \ \g\ \ $}}\ \,
M_\g \bpm \vf^a_D \\ \vf^a \epm, \qquad
M_\g \in \Sp(2r,\Z),
\end{align}
under analytic continuation along a closed path $\g$ in $\cM(\gf)$ which does not intersect the orbifold fixed point loci.  If the lift of $\g$ to the $\C^{3r}$ covering space of the orbifold is an open path with endpoints related by the action of an element $w_\g \in \cW(\gf) \subset \GL(r,\Z)$, then the associated EM duality monodromy in \eqref{N4EMmonod} is
\begin{align}\label{N4EMmonod2}
M_\g = \bpm w_\g & 0 \\ 0 & w_\g^{-T} \epm  \in \Sp(2r,\Z).
\end{align}

The $\cN=4$ sYM theory can be viewed as an $\cN=2$ theory with respect to a choice of an $\cN=2$ subalgebra of the $\cN=4$ superconformal algebra.  From this point of view, the $\cN=4$ Coulomb branch decomposes into an $\cN=2$ Coulomb branch $\cC(\gf)$ (an $r$ complex-dimensional special K\"ahler space) and an $\cN=2$ Higgs branch $\cH(\gf)$ (an $r$ quaternionic-dimensional hyperk\"ahler space) which are each subspaces of a $3r$ complex dimensional enhanced Coulomb branch \cite{Argyres:2016ccharges}.  The geometries of these Coulomb and Higgs branches are induced from the geometry of 
$\cM(\gf)$ in the obvious way, as the flat orbifolds
\begin{align}\label{N4asN2sYM}
\cC(\gf) = \C^{r} /\cW(\gf) ,
\qquad 
\cH(\gf) = \C^{2\,r} /\cW(\gf) .
\end{align}
The special K\"ahler structure of the $\cN=2$ Coulomb branch is just the restriction of the one described above for the $\cN=4$ Coulomb branch.  The hyperkahler structure of the $\cN=2$ Higgs branch can be descibed as follows.  Choose a complex structure on $\cH(\gf)$ with flat complex coordinates $\z^a_i$ for $a=1,2$ such that $(\z^1_i , \bar\z^2_i)$ transform as a doublet under the $\SU(2)_R$ isometry, i.e., so that $\z^1_i = \vf^1_i$ and $\z^2_i = \bar\vf^2_i$.  Then the K\"ahler form of $\cH(\gf)$ with respect to this complex structure is $\w^{(1,1)} = C^{ij} ( d\z^1_i \^ d\bar\z^1_j + d\z^2_i \^ d\bar\z^2_j )$ and the holomorphic 2-form made from the K\"ahler forms with respect to the other two orthogonal complex structures of $\cH(\gf)$ is $\w^{(2,0)} = C^{ij} d\z^1_i \^ d\z^2_j$.

The complex structure of $\cC(\gf)$ turns out to be very simple: as a complex space the $\cN=2$ Coulomb branch is isomorphic $\C^r$ and thus is regular, though, of course, it still has metric singularities (non-analyticities) at the orbifold fixed-point loci.  It follows that the Coulomb branch chiral ring of the $\cN=4$ sYM OPE algebra is freely generated.  In section \ref{Cstructure} we will discuss how to derive the complex structure of the CB in a systematic way.  In contrast, the complex structure of the $\cN=2$ Higgs branch is less trivial and the Higgs branch chiral ring is generically not freely generated.  We will discuss this briefly in section \ref{Cstructure} as well.

\subsection{Moduli space geometry of the gauged theories}

Upon gauging one of the discrete $\G_k$ symmetries constructed above in section \ref{construct}, the $\cN=4$ moduli space orbifold \eqref{N4sYM} will be further identified by the corresponding action of $\G_k$ on the Cartan subalgebra scalars given in \eqref{Gam2CBact}, \eqref{tGam2CBact}, \eqref{tGam3CBact}, or \eqref{GamkCBact}.  Thus the new moduli space of vacua of the discretely-gauged $\cN=4$ sYM theory with gauge algebra $\gf$ will be
\begin{align}\label{N3mod}
\cM_k(\gf) := \cM(\gf)/\G_k = [\C^{3\,r} /\cW(\gf)]/\G_k 
= \C^{3\,r} / (\cW(\gf)\rtimes\G_k).
\end{align}
The corresponding Coulomb branch in an $\cN=2$ decomposition of the moduli space will be given by
\begin{align}\label{N3CB}
\cC_k(\gf) := \cC(\gf)/\G_k = 
\C^{r} / (\cW(\gf)\rtimes\G_k),
\end{align}
and similarly for the Higgs branch:
\begin{align}\label{N3HB}
\cH_k(\gf) := \cH(\gf)/\G_k = 
\C^{2r} / (\cW(\gf)\rtimes\G_k),
\end{align}
In fact, since in most cases\footnote{The exceptions are the action of $\til C_2$ in \eqref{tGam2CBact} for $\gf=\sof(4N)$ and $\til C_3$ in \eqref{tGam3CBact} for $\gf=\sof(8)$.} the $\G_k$ generators act by multiplication by overall phases so commute with the $\cW(\gf)$ generators, the orbifold group in \eqref{N3mod} is simply a direct product $\cW(\gf)\rtimes\G_k = \cW(\gf) \times \G_k$.

We start with the case of the discrete symmetries described in section \ref{N4toN4symm} preserving the $\cN=4$ supersymmetry for $\gf=\suf(N)$, $\sof(2N)$, and $E_6$ sYM theories.  (The other gauge algebras do not have any outer automorphisms, and inner automorphisms are part of the gauge group and thus cannot be further gauged.)  In all cases, the $\G_2\simeq\Z_2$ symmetry acts by an outer automorphism of the $\gf$ on the vector multiplet, and in the $\sof(8)$ case there are also outer automorphism $\tG_3\simeq\Z_3$ and $\tG_6\simeq S_3$ symmetries.  
Gauging these symmetries effectively extends the gauge group in these theories from the original $G$ to $G \rtimes \G_2$, and similiarly for the other outer automorphism groups for the cases where Lie$(G)=\gf=\sof(8)$.  Such extensions of $G$ always exist since the semi-direct product group action is defined by the action of the outer automorphism group on $G$.

The geometry of the resulting $\cN=4$ CB is given by \eqref{N3mod}.  Since $\G_2$ acts by overall sign flips for each $(\vf_i^1, \vf_i^2,\vf_i^3) \in \C^3 \simeq \R^6$, it is clear that the action of $\G_2$ is in the center of the $\SU(4)_R \simeq \SO(6)_R$ isometry group.  This is a necessary condition for the $\G_2$ orbifolding to preserve $\cN=4$ supersymmetry on the moduli space.  This is less obvious for the $\tG_3$ and $\tG_6$ orbifold actions in the $\gf=\sof(8)$ theories, but follows because the generator $\til C_3$ preserves the $C^{ij} d\vf^a_i d\bar\vf_a^j$ hermitean metric.
 
As discussed in \cite{allm1602}, in many cases the orbifold groups entering the description of the moduli space of the gauged theory in \eqref{N3mod} are themselves Weyl groups:
\begin{align}\label{Weyl2Weyl}
\cW(\suf(3))\rtimes\G_2 &= \cW(G_2) ,\nonumber \\
\cW(\suf(4))\rtimes\G_2 &= \cW(\sof(7)) = \cW(\mathfrak{sp}(6)) ,\nonumber \\
\cW(\sof(8))\rtimes\tG_6 &= \cW(F_4) ,\\
\cW(\sof(2r))\rtimes\G_2 &= \cW(\sof(2r+1)) = \cW(\mathfrak{sp}(2r)) 
\quad\text{for}\quad r\ge1 .\nonumber
\end{align}
This means there are distinct $\cN=4$ field theories sharing identical moduli space geometries.  For example this shows that the moduli space of the $\cN=4$ sYM theory with gauge group $G_2$ is the same as the moduli space of an $\cN=4$ sYM theory with gauge group $\SU(3) \rtimes \Z_2$.

But the other cases --- namely, $\cW(A_r)\times\G_2$ for $r \ge 4$, $\cW(D_4)\rtimes\tG_3$, and $\cW(E_6)\times\G_2$ --- give new orbifold groups and thus new $\cN=4$ moduli spaces.  The main question addressed in this paper is what are the complex structures of the resulting $\cN=2$ Coulomb branch geometries \eqref{N3CB}? As we will discuss in detail in the next two sections, providing explicit constructions in section \ref{examples}, these generally give $\cN=2$ Coulomb branch geometries with complex singularities.  Thus they give examples of SCFTs with non-freely-generated Coulomb branch chiral rings.

A very similar story holds for the theories where $\G_k$ for $k>2$ is discretely gauged.  As discussed earlier, these symmetries only exist when the sYM coupling is fixed at special strong-coupling values, so the $\G_k$-gauged theories have no marginal coupling and, in particular, no weakly-coupled description in terms of gauge fields.  We know of no clear sense in which we can describe these new theories as sYM theories with extended gauge groups.  Nevertheless, the geometry of their moduli spaces is still described by \eqref{N3mod}.  The action of the $\G_k$ on the moduli space fields given in \eqref{GamkCBact} preserves an $\cN=3$ supersymmetry.  Indeed, in a complex structure in which local complex coordinates are taken to be $(\vf^1_i, \vf^2_i, \bar\vf^3_i)$, $\G_k$ acts as an overall phase rotation of all coordinates, and so commutes with the $\U(3)_R$ isometry group which acts linearly on these triplets; this is a necessary condition for $\cM_k(\gf)$ to be the moduli space of an $\cN=3$ SCFT.  And similarly to the $\G_2$ case, their orbifold groups $\cW(\gf)\rtimes \G_k$ are generally not complex reflection groups and give additional examples of $\cN=2$ Coulomb branches with complex singularities.  Again, a more systematic discussion will be presented in the next sections.

Finally, some comments about extensions of the constructions of this section to theories witht less supersymmetry will appear in section \ref{questions}.

\section{Complex structure of the CB}\label{Cstructure}

We are now ready to perform explicitly the discrete gauging described in the previous section and analyze in detail the complex structure of the $\cN=2$ CBs $\cC(\gf)$ and $\cC_k(\gf)$ defined in \eqref{N4asN2sYM} and \eqref{N3CB}.  Before turning to explicit constructions, which will be the content of the next section, we will describe the mathematical tools we are going to use in the analysis.  We will use the symbol $\cC$ with no extra label to refer to properties which apply equally to $\cC(\gf)$ and $\cC_k(\gf)$ and throughout our analysis $r$ indicates the rank of the associated conformal theory, that is $\dim_\C\cC=r$.  At the end of this section we will also present a brief discussion of the geometry of the $\cN=2$ Higgs branches, $\cH(\gf)$ and $\cH_k(\gf)$.

\subsection{General considerations}

As an affine algebraic variety, $\cC$ is defined as the common zeros of a set of polynomials in $n$ variables $(u_1,\ldots,u_n)$,
\begin{align}\label{compvar}
\cC = \{ (u_1,\ldots, u_n)\in\C^n \ |\ P_k(u_1,...,u_n)=0 \}.
\end{align}
The 4d $\cN=2$ superconformal algebra contains an $\SO(1,1)_D \times \U(1)_R$ dilatation and R-symmetry which combine to give a non-trivial holomorphic $\C^*$ action on $\cC$.  We take the $\C^*$ action to act on the affine coordinates as $\C^*: u_a \mapsto \l^{\D_a} u_a$ for $\l\in\C^*$ with definite positive scaling dimensions $(\D_1,...,\D_n)$.  Then the $P_k$ are weighted homogeneous polynomials of degree $\D_{P_k}$,
\beq\label{Homo}
P_k(\l^{\D_1}u_1,...,\l^{\D_n}u_n)=\l^{\D_{P_k}}P_k(u_1,...,u_n).
\eeq
$\cC$ is singular at $\boldsymbol{u}_0 :=(u_1^0,...,u_n^0)$ as a complex variety if and only if
\begin{align}\label{singcond}
dP_k|_{\boldsymbol{u}_0} = 0
\qquad\text{for all $k$}.
\end{align}
Note that \eqref{Homo}, \eqref{singcond}, and Euler's theorem,  $\sum_i \D_i u_i\del_{u_i}P_k=\D_{P_k}P_k$, imply $P_k(\boldsymbol{u_0})=0$.

If any $P_k$ has a single $u_i$ alone as one of its terms, then that $u_i$ can be eliminated in terms of the other $u_j$'s, and that $P_k$ can also be dropped (since it is then identically satisfied).  After eliminating all such $u_i$, either none of the remaining $u_i$ appears alone in any term of the remaining $P_k$, or all the $P_k$ are identically satisfied.  In the latter case the CB is isomorphic, as a complex variety, to $\C^r$.\footnote{This is the way the CBs of many $\cS$-class SCFTs constructed in \cite{Chacaltana:2017boe} end up being freely generated.}  The former case implies that $dP_k^0(u_i)|_{u_i=0}=0$, and thus the algebraic variety described by $P_k^0(u_i)=0$ is singular for $u_i=0$. Notice that this argument only applies to the affine coordinates which appear in the defining algebraic equations of the variety \eqref{compvar}. In the case in which all the $P_k$'s are independent of one or more of the affine coordinates, the complex singularity is not isolated and the geometry has a $\til n$ complex dimensional singular locus, where $\til n$ is equal to the number of affine coordinates which are unconstrained by the relations.  We will see below that the CBs that we are going to construct generically have non-isolated singularities of this kind. 

For our analysis it is more natural, though equivalent, to describe $\cC$ through its coordinate ring,
\beq\label{quot}
\C\left\{\cC \right\}:=\C[u_1,...,u_n]/\cI\left(\cC\right),
\eeq
where $\C[u_1,...,u_n]$ is the polynomial ring over $\C^n$, the affine space where $\cC$ can be embedded algebraically. $\cI\big(\cC\big)$ is the ideal generated by non-trivial relations identiclly satisfied by the $u_i$'s on $\cC$; that is, $\cI\big( \cC\big)$ contains all polynomials which vanish at all points on $\cC$. Drawing a connection between the two descriptions is straightforward: the $P_k(u_i)$'s in \eqref{Homo} are precisely the generators of $\cI\big(\cC\big)$. Since the $u_i$'s have definite scaling dimension, $\C[u_1,...,u_n]$ is a graded ring, and from its definition $\cI\big(\cC\big)$ is a homogeneous ideal.  Thus the coordinate ring $\C\{\cC\}$ \eqref{quot} is itself a graded ring. This will be useful in our analysis below.  With the assumption that  all CB chiral fields of the SCFT can get vevs (i.e., correspond to flat directions) consistent with their chiral ring relations, and that the CB chiral ring is reduced (i.e., has no nilpotents), then the CB coordinate ring \eqref{quot} and CB chiral ring coincide.

As discussed in the previous section, the CBs which we will construct here can be written globally as orbifolds,\footnote{For a discussion of whether this is a general property of moduli spaces of SCFTs with $\cN\geq3$, see \cite{ClassN3}.  Generically, moduli spaces of $\cN=2$ SCFTs are not orbifolds.}
\beq\label{orbi}
\cC\equiv \C^r/\G ,
\eeq
where $\G$ is a finite group, either $\cW(\gf)$ as in \eqref{N4asN2sYM} or $\cW(\gf)\rtimes \G_k$ as in \eqref{N3CB}.  For orbifolds, the coordinate ring \eqref{quot} of the CB is
\begin{align}\label{}
\C\left\{\C^r/\G \right\} = \C[z_1,\ldots,z_r]_\G := J_\G,
\end{align}
where $J_\G$ is the (graded) {\it ring of polynomial invariants} of the $\G$-action on $\C^r$.  This can be described as the coordinate ring of an affine algebraic variety as in \eqref{quot} by taking the affine coordinates $(u_1,\ldots,u_n)$ to be an algebraically independent basis of the invariant polynomials in $r$ variables of $\G$, and the ideal of defining equations, $\cI(\cC)$, to be the ideal generated by the algebraic relations identically satisfied by the $u_k(z_i)$.

If $\cI\big(\cC\big)$ is trivial, then $\C\left\{\cC\right\}\equiv\C[u_1,...,u_n]$, that is the coordinate ring is simply a polynomial ring over $\C^n$ and the associated CB chiral ring is freely generated.  Since orbifolding does not change the dimension of the CB, it also follows that $n=r$, the rank of the SCFT under consideration.  Conversely, as argued above, if $\cI\big(\cC\big)$ is not trivial, then by virtue of its $\C^*$ symmetry, $\cC$ will have a (perhaps non-isolated) complex singularity, and the associated coordinate ring and CB chiral ring are not freely generated.  So the key question is to determine whether $\cI\big(\cC\big)$ is trivial for complex orbifolds like \eqref{orbi}.

In fact, a powerful theorem by Chavalley, Shephard and Todd (CST) \cite{Shephard1954,Chevalley1955} proves that the ring of invariants of an orbifold action \eqref{orbi} is a polynomial ring if and only if $\G$ is a complex reflection group acting irreducibly on $\C^r$.  Furthermore consistency of the low energy theory on the CB under EM duality transformations requires that the group acting on $\C^r$ be crystallographic \cite{Caorsi:2018ahl}.  A full classification of crystallographic complex reflection group is given in \cite{Popov1982}.

In the orbifold CBs constructed in the last section, the orbifold group $\G$ was either the Weyl group $\G = \cW(\gf)$ for the parent $\cN=4$ sYM theory, or one of its extensions $\G = \cW(\gf)\rtimes \G_k$ for $k=2,3,4,6$ where $\G_k \simeq \Z_k$ with a specified linear action on $\C^r$.

The Weyl groups of simple Lie algebras are precisely the irreducible real crystallographic reflection groups \cite{humphreys1990coxeter}.  So by the CST theorem $\C\{\cC(\gf)\}\cong\C[u_1,...,u_r]$ is a polynomial ring with the $u_i$ a basis of the invariant polynomials in $r$ variables of $\cW(\gf)$.  The content of the CST theorem is that any such basis satisfies no further non-trivial relations, so $\cC(\gf)$ has no complex singularities and as a complex manifold is simply isomorphic to $\C^r$.  The scaling dimensions of the $u_i$ are given by the degrees of the adjoint Casimirs of $\gf$. 

For a daughter theory to have a freely-generated CB coordinate ring, CST says that $\G = \cW(\gf)\rtimes\G_k$ must be a complex reflection group.  This becomes an increasingly stringent constraint as the rank of $\gf$ increases, and so we generically expect that the daughter theory CBs will have complex singualrities.

In the case of $\cN=4$ supersymmetric daughter theories with $\G=\cW(\gf)\rtimes\G_2$ (or the other two possibilities when $\gf=\sof(8)$), low energy $\cN=4$ supersymmetry requires a complex reflection group $\G$ to actually be a real reflection group in order for the orbifold action to preserve an $\SO(6)_R$ group of isometries on the moduli space as explained above eqn.\ \eqref{Weyl2Weyl}.  Since the only real crystallographic reflection groups are Weyl groups, the only cases in which an $\cN=4$ daughter theory CB can have a freely-generated coordinate ring is if $\cW(\gf)\rtimes\G_k$ is itself another Weyl group.  All such cases are listed in \eqref{Weyl2Weyl}.  It therefore follows that the other cases --- namely, $\cW(A_r)\times\G_2$ for $r \ge 4$, $\cW(D_4)\rtimes\tG_3$, and $\cW(E_6)\times\G_2$ --- give $\cN=4$ moduli spaces whose CBs have complex singularities.

In the case of $\cN=3$ daughters with $\G=\cW(\gf)\rtimes\G_k$ for $k=3,4,6$, the question then becomes whether or not $\G$ is a complex reflection group.  We will see below that the answer is that they are generically not reflection groups, and so their CB orbifolds generically have complex singularities.

\subsection{Hilbert series of rings of polynomial invariants}

It is now time to delve into understanding how to compute $J_\G$ in a way in which we can read off its generators and the relations they satisfy to derive an explicit expression for the CB coordinate ring \eqref{quot}.  In particular, we will review a mathematical tool --- the Hilbert series of the coordinate ring --- that will be useful for describing the complex structure of $\cC(\gf)_k$ in the case where $\cW(\gf)\rtimes\G_k$ is not a complex reflection group, and so the CST theorem does not apply.  Although the Hilbert series does not give complete information on the coordinate ring, it has the advantage of being easily computable for a ring of invariants $J_\G$ of a finite group $\G$ acting on $\C^r$.  In many cases knowing the Hilbert series will allow us to explicitly compute the generators of $J_\G$ and the relations which they satisfy, and thus to reconstruct the coordinate ring of the orbifold space \eqref{quot}.

Recall that the coordinate ring $\C\{\cC\}$ of an affine algebraic variety \eqref{quot} describing a SCFT moduli space is a graded $\C$-algebra by virtue of the $\C^*$ action,
\begin{align}\label{}
\C\{\cC\}=\bigoplus_{j\geq0}\ \C\{\cC\}\big|_j ,
\end{align}
where the grading is given by the homogeneous degree of the polynomials in the $\C^r$ coordinates $(u_1,...,u_n)$. 
Its Hilbert series \cite{CommAlg}, 
\begin{align}\label{}
P_\cC = \sum_{j=0}^\infty p_j t^j ,
\end{align}
is a formal series in a variable $t$ with non-negative integer coefficients $p_j := \dim (\C\{\cC\}\big|_j)$.  That is, $p_j$ gives us the number of linearly independent homogenous polynomials of degree $j$ in $\C\{\cC\}$.  

In general for affine algebraic varieties, the Hilbert series has the form
\begin{align}\label{HPseries}
P_\cC(t) = \frac{Q(t)}{\prod_{j=1}^n(1-t^{d_j})}
\end{align}
when the affine parameters $(u_1,...,u_n)$ have scaling dimensions $(d_1,...,d_n)$.  Here $Q(t)$ is a polynomial whose form encodes properties of the ideal $\cI(\cC)$.  For example, in the case $\cC$ is a complete intersection (c.i.) whose coordinate ring is $\C[u_1,...,u_n]/\til\cI$, where $\til\cI$ is a free module generated by a set of relations $(\th_1,...,\th_m)$ of degrees $(\til d_1,...,\til d_m)$, the Hilbert series is \cite{Stanley}
\beq\label{polci}
P_{\rm{c. i.}}(t) = \frac{\prod_{i=1}^m(1-t^{\til d_i})}{\prod_{i=j}^n (1-t^{d_j})} .
\eeq
The reader can check by expanding \eqref{polci} that the coefficient of $t^k$ gives in fact the right number of independent homogenous polynomials of total degree $k$ generated by a basis of parameters of degrees $(d_1,...,d_m)$ with independent relations at degrees $(\til d_1,..., \til d_m)$.  

In the case we are interested in, where the coordinate ring is the ring of polynomial invariants of a finite linear group action, $\C\{\cC\} = J_\G$, the Hilbert series is given by Molien's formula \cite{Molien},
\beq\label{Molien}
P_{J_\G}(t)=\frac{1}{|\G|}\sum_{g\in \G}\frac{1}{\det(\I-g t)}.
\eeq
In this case the Hilbert series is sometimes called the Molien series.  \eqref{Molien} has the advantage that its series expansion in $t$ is easily computable given a $\G$ action on $\C^n$.  The closed-form rational expression \eqref{HPseries} for $P_{J_\G}$ is not so readily computable, however, once the order $|\G|$ of the group gets large.

Often, in our analysis below, it turns out to be more convenient to consider the orbifold action of $\G_k$ on $\cC(\gf)$ rather than the $\cW(\gf)\rtimes\G_k$ action on $\C^r$ \eqref{N3CB}.  Call $(u_1,...,u_r)$ the basis of the coordinate ring of $\cC(\gf)$, the $u_i$ are themselves graded by their scaling dimensions $\D_i$ (or alternatively by their $\U(1)_R$ charges).  In such cases we can refine the Hilbert series by keeping track of this extra grading to distinguish not just the overall degree of the homogeneous polynomials but also their individual degrees in the $u_i$'s.  Since the grading of $\cC(\gf)$ obviously depends on $\gf$, to avoid ambiguity, we will denote the ring of invariants of $\G_k$ on $\cC(\gf)$ as $J^\gf_{\G_k}$.  $J^\gf_{\G_k}$ has the direct sum decomposition, 
\beq\label{refined}
J^\gf_{\G_k}=\bigoplus_{i_1,...,i_\ell\geq0}J^\gf_{\G_k}\big|_{i_1,...,i_\ell} ,
\eeq
where $J^\gf_{\G_k}\big|_{i_1,...,i_\ell}$ only contains homogeneous polynomials with degree $i_j$ in the $u_j$'s with scaling dimension $\D_{i_j}$. Notice that $\ell\leq r$ as some of the $u_i$ might have the same scaling dimension.  The dimension of the $J^\gf_{\G_k}\big|_{i_1,...,i_\ell}$'s is computed from the refined Molien series \cite{Benvenuti:2006qr,Feng:2007ur}, 
\beq\label{Molref}
P_{J^\gf_{\G_k}}(t_1,...,t_\ell)=\frac{1}{|\G_k|}\sum_{g\in \G_k}\frac{1}{\det(\I-g\ {\rm diag}(t_1,...,t_\ell))}.
\eeq
Just as with the Hilbert series, it is a formal power series in $(t_1,\ldots,t_\ell)$, and the coefficient of the $t_1^{i_1} \cdots t_\ell^{i_\ell}$ term is $\dim(J^\gf_{\G_k}\big|_{i_1,...,i_\ell})$.

\subsection{Counting generators and relations}

The expressions \eqref{HPseries} and \eqref{polci} for the Hilbert series clearly indicate that the Hilbert series encodes information about the generators of the coordinate ring and their degrees, as well as of the generators and degrees of the ideal of equations or ``relations" defining $\cC$.  This data appears in the form of the expressions for the Hilbert series written as rational functions, but, for the Molien series computed from \eqref{Molien} or \eqref{Molref} due to computer power limitations for large-order groups, we generally only have access to some finite number of leading terms of the Hilbert series as a series in $t$.  So to extract information about generators and relations we need a way to ``invert" expressions like \eqref{HPseries} given only partial information about the right side of the expression.

The ``plethystic logarithm" or the ``inverse of the plethystic exponential" \cite{PleExp,Labastida:2001ts} of the Molien series is such a tool.  It is defined as
\beq\label{plet}
\cF_\G(t) := PE^{-1}\left(P_{J_\G}(t)\right) =
\sum_{m=1}^\infty\frac{\m(m)}{m}
\log\left(P_{J_\G}(t^m)\right) ,
\eeq
where $\m(m)$ is the M\"obius function,
\beq
\m(m) =
\begin{cases}
0 & m\ \text{has one or more repeated prime factors}\\
1 & m=1\\
(-1)^n& m\ \text{is a product of $n$ distinct primes}
\end{cases} .
\eeq
The resulting function $\cF_\G(t)$ is another formal power series in $t$ with, not necessarily positive, integer coefficients, and is easily computable using the power series expansion of the logarithm around 1.  It is essentially designed to extract the counting of generators and relations in the form
\beq\label{FormSer}
\cF_\G(t) \overset{?}{=}  \underbrace{\sum_kc^{+}_kt^k}_\textrm{generators}-\overbrace{\sum_{k'}c^{-}_{k'}t^{k'}}^\textrm{relations},\qquad c_k^+,c_{k'}^-\in\N ,
\eeq
where the positive coefficients in $\cF_\G(t)$, $c_k^+$, count the number of generators of degree $k$ while negative coefficients $c_{k'}^-$ count the number of relations at degree $k'$.  The question mark in \eqref{FormSer} indicates that it is not true in general, as we will discuss below.  But it is easy to see that it works precisely in the complete intersection case: the reader can check that if the Molien series has the form \eqref{polci}, then its plethystic logarithm is
\beq
\cF(t)=PE^{-1}\left(\frac{\prod_{i=1}^{m}(1-t^{\til d_i})}{\prod_{j=1}^n (1-t^{d_j})}\right)=t^{d_1}+...+t^{d_n}-t^{\til d_1}-...-t^{\til d_m} .
\eeq
So if the orbifold is a complete intersection --- that is, $\cI(\cC)$ is a free module of rank $m$ --- then the plethystic logarithm series truncates to the polynomial \eqref{FormSer}.  

Finally \eqref{plet} generalizes straighforwardly to the refined Molien series,
\begin{align}\label{refinedPL}
\cF^\gf_{\G_k}(t_1,...,t_\ell) :=
PE^{-1}\left(P_{J^\gf_{\G_k}}(t_1,...,t_\ell)\right)=
\sum_{m=1}^\infty\frac{\mu(m)}{m}\log\left(P_{J^\gf_{\G_k}}(t^m_1,...,t^m_\ell)\right) ,
\end{align}
where we again use the explicit label $\gf$ to keep track of the fact that the orbifold is on $\cC(\gf)$ and not $\C^r$. $\cF^\gf_{\G_k}(t_1,...,t_\ell)$ is a formal series in the $t_i$'s and $c_{i_1,...,i_\ell}^+$ count the number of generators of degree $i_i$ in the $u_j$'s with scaling dimension $\D_{i_i}$ while negative coefficients $c_{i'_1,...,i'_\ell}^-$ count the number of relations at degree $i'_i$ in the $u_j$'s with scaling dimension $\D_{i'_i}$. And as before $\ell\leq r$ as some of the $u_i$'s might have the same scaling dimension.

How much information about the generators and relations of a coordinate ring can be extracted from its Hilbert series or its plethystic logarithm?  

Note first that $\cI(\cC)$ in \eqref{quot} may itself not be a freely-generated module.  That is, its generators (which we think of as describing relations among the affine parameters defining $\cC$) may themselves satisfy non-trivial relations.  The existence of relations among relations, or ``syzygies", is very often the case unless the rank of $\cI(\cC)$ is 1 --- that is, $\cI(\cC)$ is generated by a single element --- in which case $\cI(\cC)$ is obviously free.  The existence of syzygies means that the resulting CB is not a complete intersection, and the numerator $Q(t)$ of the Hilbert series in \eqref{HPseries} need not have the simple factorized form \eqref{polci}.  In such a case the plethystic logarithm is no longer a polynomial, but is instead is an infinite series, and so the simple interpretation \eqref{FormSer} of its coefficients cannot be true.

It is tempting, nevertheless, to interpret just the leading terms of the plethystic logarithm as in \eqref{FormSer}.  The idea is that if the generators appear at low degrees they will contribute to the leading positive-coefficient terms of the series, while relations of generators will typically be at higher degree and will contribute to the next set of negative-coefficient terms, and then relations-among-relations would be at still higher degrees and so on.  Indeed, \cite{Benvenuti:2006qr,Feng:2007ur} conjecture that ``the plethystic logarithm of the Molien series is a generating series for the relations and syzygies of the variety." 

However, this conjecture cannot work in all generality.  Indeed, it is easy to construct simple counter-examples where it fails.  For instance, examples 3.8 and 3.9 in \cite{Stanley} give instances where the Hilbert series fails to encode the generators and relations in the way described above.  Example 3.8 is a case of a complete intersection whose Hilbert series has the same form as that of a freely-generated coordinate ring, and example 3.9 is that of a non-complete-intersection variety whose Hilbert series nevertheless has the form \eqref{polci} expected of a complete intersection.   The basic reason that these examples violate the conjecture is that there are ``unexpected" cancellations between factors in the numerator and denominator of the Hilbert series \eqref{HPseries}.  This can happen when the degree of a relation happens to be the same as that of an affine parameter in the coordinate ring, or if the degree of a syzygy happens to coincide with that of a relation, etc.  As the rank of $\cC$ increases, such accidental cancellations become more likely, but, at least for low-rank examples, one might expect that the plethystic logarithm will accurately capture the degrees and counting of generators and relations.  Furthermore, by using the refined Molien series and its plethystic logarithm, \eqref{Molref} and \eqref{refinedPL}, many accidental cancellations can be resolved as the factors of $t^k$ with cancelling coefficients now may be different monomials of total degree $k$ in the $t_j$'s.  

Indeed, the plethystic logarithm interpretation \eqref{FormSer} works surprisingly well (as we will also see below) in reproducing generators and relations of known orbifolds \cite{Benvenuti:2006qr, Feng:2007ur}.  In the generic case, though, where the series for $\cF_\G(t)$ no longer truncates, a certain amount of guessing is involved in understanding how to precisely interpret the coefficients in the expansion \eqref{FormSer}.  We will come back to this point in specific examples in the next section.

\subsection{Comments on Higgs branch complex geometry}

All of the theories that we will analyze below have $\cN\geq3$, thus the CB is part of a larger moduli space and in particular all theories have a non-trivial Higgs branch. We will not give a systematic analysis of the Higgs branch complex geometry nor its chiral ring, but it will be useful to outline a few facts about their complex geometry. $\cH(\gf)$ and $\cH_k(\gf)$ indicate the Higgs branches of the parent \eqref{N4asN2sYM} and daughter \eqref{N3HB} theory respectively. We will use $\cH$ to refer to properties which apply in both cases, for example $\dim_\C\cH=2r$.  Since the Higgs branch geometries are orbifolds by a finite group $\G$ in all the cases analyzed in this paper, their coordinate rings are isomorphic to rings of $\G$-invariant polynomials and we can apply the same reasoning and techniques outlined above for the CB to this case. 

But, unlike the CB case, the HB coordinate ring is generically not freely-generated even when $\G$ is a complex reflection group.  The Higgs branch orbifold is $\cH \sim \C^{2r}/\G$ where the action of $\G$ on $\C^{2r}$ is given by the direct sum of two copies of its irreducible action on $\C^r$ considered previously.  Calling $\r_r(\G)$ the $r$-dimensional representation which acts irreducibly on $\C^r$, we take $\r_{2r}(\G) := \r_r(\G) \oplus \bar\r_r(\G)$, where $\bar\r_r(\G)$ is the complex conjugate representation.\footnote{Clearly $\r(\cW(\gf)) \cong \bar\r(\cW(\gf))$ since Weyl groups are real reflection groups, thus reproducing the action \eqref{N4EMmonod2} in the $\cN=4$ case.}  By construction, $\r_{2r}(\G)$ does not act irreducibly on $\C^{2r}$ and thus Chevalley-Shephard-Todd theorem no longer applies.  It follows that coordinate rings of Higgs branches are generically not freely generated.

To see this explicitly, write the Higgs branch coordinate ring as 
$\C\{\cH\} = \C[z^1_1, \ldots, z^1_r,$ $z^2_1, \ldots, z^2_r]_\G$.  Even though the action of each Weyl group element splits as a direct sum of actions on $\C^r \times \C^r$, in addition to the $u_k$ invariants built from just $z^1_i$'s and similar $v_k$ invariants built from just $z^2_i$'s, there will now be many more invariant polynomials of the same or lower degrees containing mixtures of the $z^1_i$'s and the $z^2_i$'s.  Furthermore, since the total dimension of the Higgs branch is $2r$, these invariants cannot all be algebraically independent, so the chiral ring will have non-trivial relations, and the complex structure of $\cH(\gf)$ is therefore not regular.  This occurs even in the simplest example, where $r=1$ and $\cW(\suf(2)) = \Z_2$ is generated by $-I_2$:  then $\cH(\suf(2)) = \C^2/\Z_2 = \C[u, v, w]/\vev{u v - w^2}$ where $u=(z^1_1)^2$, $v=(z^2_1)^2$, and $w = z^1_1 z^2_1$.  

It is worth pointing out that in the case of the Higgs branch we can use a Molien series refined by the natural grading for the orbifold of $\C^{2r}$ provided by the $\U(2)_R$ isometry.  Choose a parametrization for $\C^{3r}=\C^3\otimes\C^r$ as $(\bz^a)$, where $a=1,2,3$ and $\bz^a\in\C^r$.  We choose $\bz^a$ with $a=2,3$ as the $\C^{2r}$ which gives the HB. The $\U(3)_R \subset \SU(4)_R$ acts as: $\r_{3r}(\U(3)_R) := \U(3)_R \otimes \I_{r\times r}$ on $(\bz^1, \bz^2, \bar\bz^3)$, which implies that coordinates $\bz^2$ and $\bz^3$ carry different charges under a $\U(1)^3_R$ maximal torus of $\U(3)_R$.  This is also the reason why the appropriate $\r_{2r}(\G)$ which commutes with the $R$-symmetry involves a direct sum of $\r_r(\G)$ and $\bar\r_r(\G)$; for more details see \cite{ClassN3}.  We will not perform any detailed HB calculations here.

\section{Examples}\label{examples}

We will first analyze the gauging of the outer automorphism group which preserves all the $\cN=4$ supersymmetry and then move to few examples of discrete gauging which only preserve 3 of the 4 supercharges. 

\subsection{$\cN=4$ theories with regular CBs}

Let us start with studying the set of $\cN=4$ theories whose CB remains freely generated even after the gauging of their outer automorphism group.  As mentioned in passing in the previous section, these theories represent a somewhat special set.  In fact by the Chavalley-Shephard-Todd theorem \cite{Shephard1954, Chevalley1955}, for the CB to be freely generated after gauging the outer automorphism symmetry, $\cW(\gf)\rtimes \Out(\gf)$ has to be a complex reflection group.  But by construction $\cW(\gf)\rtimes\Out(\gf)$ is real and crystallographic, thus it has to be a Weyl group of another Lie algebra, $\cW(\gf)\rtimes\Out(\gf) \cong \cW(\gf')$.  This has interesting implications.  The moduli space of $\cN=4$ theories are completely specified by their orbifold action \eqref{N4sYM} thus $\cM_{\rm{Out}}(\gf)\equiv\cM(\gf')$.  From the $\cN=2$ perspective this implies that the $\Out(\gf)$ discretely gauged $\cN=4$ theory with Lie algebra $\gf$ and the $\cN=4$ theory with Lie algebra $\gf'$ have not only isomorphic CBs but also identical Higgs branches and extended CBs.  It is common for $\cN=2$ to have isomorphic subcomponents of their moduli spaces, but to our knowledge this is the first example of two different theories which have identical moduli spaces.  Since discrete gauging does not change the value of the central charges, the two theories share their moduli spaces but can be distinguished by their different central charges.

\subsubsection{$\SU(3)$ $\to$ $G_2$}

Let us start with a very simple example at rank 2 and consider
\beq\label{su3}
\C^2/\big(\cW(\suf(3))\times\G_2\big)\cong\C^2/\cW(G_2).
\eeq
We will check momentarily by going through the computation of the Molien series and its plethystic logarithm, that the two orbifolds have the same coordinate rings. 

The irreducible action of $\cW(\suf(3))$ is the two dimensional representation of $S_3$, the symmetric group of degree 3 while the $\G_2\cong\Z_2$ is chosen to be the Chevalley involution defined in \eqref{Chevinvol}.  Thus generators of the orbifold action in \eqref{su3} can be chosen to be
\beq\label{gensu3}
M_1=\bpm0&1\\1&0\epm , \quad
M_2=\bpm1&0\\-1&-1\epm ,
\quad\text{and}\quad 
\G_2=\bpm-1&0\\0&-1\epm ,
\eeq 
which generate a group of order 16.  Using this action on $\C^2$ we compute \eqref{Molien} explicitly in this particular case to find
\beq\label{MsA3Z2}
P_{J_{\G_2}^{\suf(3)}}(t)=\frac{1}{(1-t^2)(1-t^6)} .
\eeq
Since \eqref{MsA3Z2} is already in a factorized form of the type in \eqref{polci} it is clear that the coordinate ring of \eqref{su3} is freely generated with generators of dimension 2 and 6.  In fact computing its plethystic logarithm we obtain,
\beq
\cF^{\suf(3)}_{\G_2}(t)=PE^{-1}\left(\frac{1}{(1-t^2)(1-t^6)}\right)=t^2+t^6.
\eeq
In terms of $(u_1,u_2)$, the CB parameters of $\cC(\suf(3))$ which have scaling dimension 2 and 3 respectively, there is a unique way of generating the two generators that we need.  Thus the two coordinates of the CB of the daughter theory are readily identified as $(\tu_1=u_1, \tu_2=u_2^3)$. 

Since the isomorphism between $\cW(\suf(3))\times \Z_2$ and $\cW(G_2)$ can be checked explicitly, it follows that the Higgs branch geometries $\cH_{2}(\suf(3))$ and $\cH(G_2)$ also coincide. It would be very interesting to check, in this low rank case, what other quantities among these two theories also match. A natural place to start is to compute the index of both theories.  

Note, however, that the S-duality groups of the two theories do not match.  The S-duality group of the $\suf(3)$ sYM theory is $\G_0(3) \subset \SL(2,\Z)$ which, as computed at the end of section \ref{N4toN3symm} has only a $\Z_2$ and a $\Z_6$ cyclic subgroup.  Discretely gauging the $\Z_2$ outer automorphism reduces the S-duality group from a subgroup of $\SL(2,\Z)$ to one of $\PSL(2,\Z)$, which reduces the cyclic subgroups to $\Z_3$ alone.  By contrast, the S-duality group of the $\gf=G_2$ sYM theory is $H_{\sqrt3} \subset \PSL(2,\R)$ whose cyclic subgroups are $\Z_2$ and $\Z_6$ (see footnotes \ref{nslg} and \ref{nslSduality}).

\subsubsection{$\SU(4)$ $\to$ $\SO(7)$}

The discussion here is very similar to the previous one. Here we will consider the orbifold
\beq\label{su4Z2}
\C^2/\big(\cW(\suf(4))\times\G_2\big)
\cong
\C^2/\cW(\sof(7)),
\eeq
and again use the computation of the Molien series and its plethystic logarithm as an extra check that the two orbifolds are isomorphic.

The irreducible action of $\cW(\suf(4))$ is the three dimensional representation of $S_4$, the symmetric group of degree 4, while the $\G_2\cong\Z_2$ is again the Chevalley involution \eqref{Chevinvol}.  It is straight forward to generate $\cW(\suf(4))\times\Z_2$ which is a group of order 64.  Having the explicit action on $\C^3$ the Molien series takes the form:
\beq\label{MSA3Z2}
P_{J_{\G_2}^{\suf(4)}}(t)=\frac{1}{(1-t^2)(1-t^4)(1-t^6)}
\eeq
Again \eqref{MSA3Z2} is in a factorized form and it is clear that the coordinate ring of \eqref{su4Z2} is freely generated with generators of dimensions 2, 4, and 6.  We will not repeat the computation of the plethystic logarithm in this case. The three coordinates of the $\cC(\suf(4))$ parent theory $(u_1,u_2,u_3)$ have scaling dimensions 2, 3, and 4 respectively.  Then the coordinates of $\cC_2(\suf(4))$ are readily identified as $(\tu_1=u_1, \tu_2=u_2^2, \tu_3=u_3)$. 

Again the Higgs branch geometries $\cH_{2}(\suf(4))$ and $\cH(\sof(7))$ also coincide and these two theories provide another explicit example of two theories with identical moduli space but different central charges and local dynamics. 

\subsubsection{$\SO(8)$ $\to$ $F_4$}\label{so8tof4}

A more interesting and somewhat surprising case, is to study the CB of the $\cN=4$ $\sof(8)$ theory with its full $S_3$ outer automorphism group gauged,
\beq\label{so8s3}
\C^4/(\cW(\sof(8))\rtimes \tG_6),
\eeq
where $\tG_6\cong S_3$.  The reader might think that because $\tG_6$ is a non-abelian finite group, the gauging would drastically change the complex structure of the initial CB and that the result of the \eqref{so8s3} must have complex singularity.  This expectation turns out to be wrong, and in fact we will momentarily see that the Molien series of this orbifold is consistent with a freely generated coordinate ring.  A posteriori the result is obvious as $\cW(\sof(8))\rtimes \tG_6 \cong \cW(F_4)$.

In order to compute the Molien series we need an explicit description of the 4 dimensional irreducible representation of $\cW(\sof(8))\cong S_4\rtimes(\Z_2)^3$ and $\tG_6$.  The former is given by considering the permutations of the four simple roots of $\sof(8)$ together with all possible sign flips of an even number of simple roots.  The latter is generated by \eqref{tGam2CBact} and \eqref{tGam3CBact}.  Then we compute the Molien series to be
\beq\label{MSD4S3}
P_{J_{\tG_6}^{\sof(8)}}(t_1,t_2,t_3)=\frac{1}{(1-t^2)(1-t^6)(1-t^8)(1-t^{12})},
\eeq
which gives a factorized answer compatible with a freely generated ring with generators of degrees 2, 6, 8, and 12.  Those are precisely the degrees of the adjoint Casimirs of $F_4$, giving a consistent picture.  As in both examples above, the $\cN=4$ theories with $\sof(8)\rtimes\tG_6$ and $F_4$ gauge algebras have identical moduli spaces but different central charges.

The information which can be extracted from \eqref{MSD4S3} isn't enough to determine the parametrization of the CB of the daughter theory in terms of the CB parameters of the parent theory. $\cC(\sof(8))$ is parametrized by four coordinate $(u_1,u_2,u'_2,u_3)$ with scaling dimensions 2, 4, 4, and 6 respectively, and there are multiple way to combine the $u$'s to get the dimension of the Casimirs of $F_4$.  The refined Molien series could help us to track exactly how the parameters of scaling dimension 8 and 12 are written in terms of the original ones.  To work out the refined Molien series, we would need the action of $\tG_6$ on $\cC(\sof(8))$ which involves a non-trivial calculation involving computing how the generators in \eqref{tGam2CBact} and \eqref{tGam3CBact} act on the invariant Casimirs of $\sof(8)$.  We will not perform this calculation here.

\subsection{$\cN=4$ theories with CB complex singulartities}

So far all $\cN=4$ theories we constructed have freely-generated CB chiral rings.  Let us now turn to the ones which develop complex singularities under the discrete gauging operation.  This is generic for $\Z_2$ gauging of $\cN=4$ $\suf(N)$ theories with $N>4$.  Thus let see how that works in the simplest case.

\subsubsection{$\Z_2$ gauging of $\SU(5)$}

The simplest example of an $\cN=4$ theory with a singular CB appears at rank 4 where the daughter theory's CB is
\beq\label{su5z2}
\C^4/\big(\cW(\suf(5))\times\G_2\big) .
\eeq
This will be the first example where we see how the computation of the Molien series and its plethystic logarithm gives us enough information to write the orbifold in a closed algebraic form.

The irreducible action of $\cW(\suf(5))$ is the four dimensional representation of $S_5$ and the $\G_2$ is again the Chevalley involution \eqref{Chevinvol}.  $\cW(\suf(5))\times\G_2$ is order 240 and the Molien series in this case is
\beq\label{MSA4Z2}
P_{J_{\G_2}^{\suf(5)}}(t)=\frac{1+t^8}{1-t^2-t^4+t^8+t^{14}-t^{18}-t^{20}+t^{22}} .
\eeq
From this expression it is not immediately obvious what the coordinate ring of \eqref{su5z2} is.  Taking the plethystic logarithm of \eqref{MSA4Z2} gives
\beq\label{gfsu5z2}
\cF^{\suf(5)}_{\G_2}(t)
=PE^{-1}\left(P_{J_{\G_2}^{\suf(5)}}\right)
=t^2+t^4+t^6+t^8+t^{10}-t^{16},
\eeq
which indicates that the coordinate ring is not freely generated but the orbifold \eqref{su5z2} is given by a hypersurface in affine $\C^5$.  The information extracted from the generating function above is not enough to specify the CB parameters of the daughter theory in terms of the parent ones.  But the refined Molien series can help us in this case.

As discussed in section \ref{N4toN4symm}, $\G_2$ acts on the vector multiplet scalars by an overall sign change \eqref{Gam2CBact}.  Call $(u_1,u_2,u_3,u_4)$ the CB coordinates of $\cC(\suf(5))$ with scaling dimensions (2,3,4,5), respectively.  Then $u_{1,3}$ must be even functions of the vector multiplet scalars, while $u_{2,4}$ must be odd ones, so
\beq
\G_2\quad:\quad
\bpm u_1\\ u_2\\ u_3\\ u_4 \epm
\quad\to\quad
\bpm u_1\\ -u_2\\ u_3\\ -u_4 \epm .
\eeq
Then the refined Molien series is readily computed to be
\beq
P_{J_{\G_2}^{\suf(5)}}(t_1,t_2,t_3,t_4)=\frac{1+t_2t_4}{(1-t_1)(1-t_2^2)(1-t_3)(1-t_4^2)},
\eeq
which is not obviously in a factorized form like \eqref{polci}. Taking its plethystic logarithm gives
\beq\label{gfsu5z2s}
\cF^{\suf(5)}_{\G_2}(t_1,t_2,t_3,t_4)
=PE^{-1}\left(P_{J_{\G_2}^{\suf(5)}}\right)
=t_1+t_2^2+t_2t_4+t_3+t_4^2-t_2^2t_4^2 ,
\eeq
which gives explicitly the complex structure of the daughter theory in terms of the parent one:
\beq\label{CBsu5z2}
\C\{\cC_2(\suf(5))\}=
\C[\tu_1,\tu_2,\tu_3,\tu_4,\tu_5]/
\vev{\tu_2\tu_4 - \tu_5^2} .
\eeq
Here $\tu_i$'s parametrize the CB of the daughter theory in terms of the CB parameters of the parent theory $u_i$'s via
\begin{align}\label{}
\tu_1=u_1, \qquad
\tu_2=u_2^2, \qquad
\tu_3=u_3, \qquad
\tu_4=u_4^2, \qquad\text{and}\qquad 
\tu_5=u_2u_4.
\end{align}
Thanks to the information provided by \eqref{gfsu5z2s} it is now easy to understand why \eqref{gfsu5z2s} was not in a factorized form.  In fact we would expect a ($1-t_1^2t_2^2)$ factor in the denominator from the relation among the generators.  But because of the $\tu_5$ generator, a $(1-t_1t_2)$ factor is also present in the denominator which partially cancels against it.  This example gives a taste of the type of cancelation which can take place in the Molien series, though in this particular case it did not lead to any loss of information about the coordinate ring.

The coordinate ring \eqref{CBsu5z2} implies that the resultant CB, as a complex variety, is a hypersurface in $\C^5$:
\beq
\cC_2(\suf(5)):=\Big\{ (\tu_1,\tu_2,\tu_3,\tu_4,\tu_5) \in \C^5 \big| \tu_2\tu_4-\tu_5^2=0 \Big\}.
\eeq
It is also interesting to look at the complex singularities of this space. Notice that the algebraic relation involves neither $\tu_1$ nor $\tu_3$.  It follows that $\cC_2(\suf(5))$ does not have an isolated complex singularity, but rather an entire two dimensional locus of complex singularities, $\cV_{\cC_2(\suf(5))}$, spanned by $\tu_1$ and $\tu_3$:
\beq
\cV_{\cC_2(\suf(5))} =
\Big\{ (\tu_1,\tu_2,\tu_3,\tu_4,\tu_5) \in \C^5
\Big| \tu_2=\tu_4=\tu_5=0 \Big\}.
\eeq

\subsubsection{$\Z_3$ gauging of $\SO(8)$}

Another somewhat surprising result is given by the
\beq\label{so8Z3}
\C^4/(\cW(\sof(8))\rtimes\tG_3)
\eeq
orbifold.  It turns out to be singular complex variety, despite the fact that $\tG_3\subset\tG_6$ and we saw above that $\tG_6$ gave rise to a non-singular complex variety. 

We already discussed how to generate $\cW(\sof(8))$.  $\tG_3$ is generated by \eqref{tGam2CBact}.  The semi-direct product of the two generates a finite group of order 576.  The Molien series is readily computed to be
\beq\label{MSD4Z3}
P_{J_{\G_3}^{\sof(8)}}(t)
=\frac{1-t^4+t^8}{(1-t^2)^4(1-t^2+t^4)(1+2t^2+2t^4+t^6)^2},
\eeq
which suggests that the coordinate ring of \eqref{so8Z3} is not freely generated.  The plethystic logarithm is
\beq\label{gfso8z3}
\cF^{\sof(8)}_{\G_3}(t)
=PE^{-1}\left(P_{J_{\G_3}^{\sof(8)}}\right)
=t^2+t^6+t^8+2t^{12}-t^{24},
\eeq
which confirms our initial guess.  More specifically, $\cC_{\til{\G}_3}(\sof(8))$ can be written as a complex variety as a hypersurface in $\C^5$:
\beq
\cC_{3}(\sof(8)):=\Big\{(\tu_1,\tu_2,\tu_3,\tu_4,\tu_5)\in\C^5\big|\tu_2\tu_3-\tu_5^3=0\Big\}
\eeq
where 
\begin{align}\label{}
\tu_1=u_1, \qquad
\tu_2=u_2^3, \qquad
\tu_3={u'_2}^3, \qquad
\tu_4=u_4, \qquad\text{and}\qquad 
\tu_5=u_2u'_2,
\end{align}
parametrize the CB of the daughter theory in terms of the CB $u_i$ parameters of the parent theory which were introduced in section \ref{so8tof4}. 

The equation above again implies that the CB of the daughter theory is a hypersurface in $\C^5$:
\beq
\cC_3(\sof(8)) :=
\Big\{ (\tu_1,\tu_2,\tu_3,\tu_4,\tu_5) \in \C^5
\big| \tu_2\tu_3-\tu_5^3 = 0 \Big\}.
\eeq
As in the previous example, the relations don't involve all of the coordinates of $\cC_3(\sof(8))$ and thus the singular locus is again two dimensional:
\beq
\cV_{\cC_3(\sof(8))} =
\Big\{ (\tu_1,\tu_2,\tu_3,\tu_4,\tu_5) \in \C^5
\Big| \tu_2=\tu_3=\tu_5=0 \Big\}.
\eeq

\subsection{$\cN=3$ theories with CB complex singularities}

Let us now work out some examples in which we break $\cN=4\to\cN=3$.  As we reviewed in some detail in section \ref{N4toN3symm}, it is well-known \cite{Goddard:1976qe,Aharony:2013hda} that the S-duality group of $\cN=4$ theories is not $\SL(2,\Z)$ in all cases and in particular its form depends on the global form, $G$, of the gauge group and not simply on its Lie algebra, $\gf$.  Here we will analyze the $G=[\SU(4)/\Z_2]_+$ $\cN=4$ theory with the specific choice of self-dual line operator spectrum \cite{Aharony:2013hda}.  The S-duality group of this theory is the full $\SL(2,\Z)$ group which contains $\Z_3$, $\Z_4$, and $\Z_6$ cyclic subgroups.   Thus we can gauge either a $\G_3$, a $\G_4$, or a $\G_6$ discrete symmetry of this theory to obtain different daughter $\cN=3$ theories.  

We will also consider the $\gf=\suf(5)$ sYM theory whose S-duality group contains only a $\Z_4$ subgroup which can be gauged.  (The two possible global forms of the gauge group, $\SU(5)$ and $\SU(5)/\Z_5$, as well as all their possible choices of line operator spectra are all exchanged by S-duality transformations, so are all part of the same theory.)

Any $\G_k$ analyzed in this section can be written explicitly in terms of $\SU(4)_R\times SL(2,\Z)$ transformations and thus its action on the CB of the parent theory can be readily obtained.  For this reason we will consider directly the refined Molien series.

\subsubsection{$\G_3$ gauging of the $[\SU(4)/\Z_2]_+$ theory}

Let's start from the simplest case and gauge $\G_3\cong\Z_3$ of the $\cN=4$ $[\SU(4)/\Z_2]_+$ theory, to find the CB
\beq
\cC_3(\suf(4)) = \cC(\suf(4))/\G_3.
\eeq
The action of the $C_3$ generator of $\G_3$ on the $\cC(\suf(4))$ coordinates $(u_1,u_2,u_3)$ of dimensions $(2,3,4)$, respectively, is 
\beq
C_3 =\bpm e^{4\pi i/3}&0&\\ 0&1&0\\ 0&0&e^{2\pi i/3}\epm ,
\eeq
as follows easily from its action \eqref{GamkCBact} on the adjoint vector multiplet scalars.

Since the three CB parameters of $\cC(\suf(4))$ have different scaling dimensions, we can use the $\U(1)_R$ grading to refine the Molien series and obtain the explicit dependence of the generators of the resulting CB in terms of $(u_1,u_2,u_3)$ as
\beq\label{MSZ3}
P_{J_{\G_3}^{\suf(4)}}(t_1,t_2,t_3)
=\frac{1+t_1t_3(1+t_1t_3)}{(1-t_1^3)(1-t_2^2)(1-t_3^3)}.
\eeq
\eqref{MSZ3} is not fully factorized. Computing its plethystic logarithm gives
\beq
\cF^{\suf(4)}_{\G_3}(t_1,t_2,t_3)
=PE^{-1}\left(P_{J_{\G_3}^{\suf(4)}}\right)
=t_1^3+t_2^2+t_1t_3+t_3^3-t_1^3t_3^3 ,
\eeq
which can be readily converted into an explicit expression for the coordinate ring of the daughter theory's CB: 
\beq\label{su4z3CR}
\C\{\cC_3(\suf(4))\}
=\C[\tu_1,\tu_2, \tu_3, \tu_4]/\langle \tu_1\tu_3-\tu_4^3\rangle ,
\eeq
where 
\begin{align}\label{}
\tu_1=u_1^3, \qquad
\tu_2=u_2, \qquad
\tu_3=u_3^3 \qquad \text{and} \qquad
\tu_4=u_1u_3.
\end{align}
So the generators of the daughter CB have scaling dimensions 6, 4, 12, and 6 respectively.  

It is worth noting that the daughter theory does not have a CB parameter with scaling dimension 2 while all the previous analyzed cases with $\cN=4$ supersymmetry, including the ones with complex singularities, did.  This is a prediction of superconformal representation theory.  Since we interpret the CB parameters as vevs of operators existing at the conformal point, the existence of a $u^*$ with $\D(u^*)=2$ implies the existence of a CB operator with $\U(1)_R$ charge 2.   Since the $\cN=4$ stress-energy tensor multiplet contains such an operator, it must occur in any $\cN=4$ SCFT.  But if it occured in an $\cN=3$ SCFT, then one of its $\cN=3$ superconformal descendants would be an additional conserved supercurrent, so the theory would actually have an $\cN=4$ supersymmetry \cite{ae1512}.  It follows that in a genuinely $\cN=3$ theory we never expect a CB parameter to have scaling dimension 2.  Our results are perfectly consistent with such expectations.  

This is closely related to the fact that $\cN=4$ SCFTs all have exactly marginal operators while genuinely $\cN=3$ theories do not, since a superconformal descendant of a dimension-2 CB operator gives an exactly marginal deformation \cite{Green:2010da}.  This is also consistent with the fact that the discrete symmetries we found in section \ref{symms} that only commuted with an $\cN=3$ subalgebra of the $\cN=4$ symmetry were also the ones which only occured at fixed values of the gauge coupling. 

The coordinate ring \eqref{su4z3CR} again implies that the resulting CB, as a complex variety, is a hypersurface in $\C^4$:
\beq
\cC_3(\suf(4)):=\Big\{(\tu_1,\tu_2,\tu_3,\tu_4)\in\C^4\big|\tu_1\tu_3-\tu_4^3=0\Big\}.
\eeq
In this case the relations involve all of the coordinates of $\cC_3(\suf(4))$ but one. Thus the space has a one complex dimensional singular locus:
\beq
\cV_{\cC_3(\suf(4))} =\Big\{(\tu_1,\tu_2,\tu_3,\tu_4)\in\C^4\Big|\tu_1=\tu_3=\tu_4=0\Big\}.
\eeq

\subsubsection{$\G_4$ gauging of the $[\SU(4)/\Z_2]_+$ theory}

We can repeat the analysis for $\G_4$ where
\beq
\cC_4(\suf(4)) =\cC(\suf(4))/\G_4
\eeq 
and the $\G_4$ action on $\cC(\suf(4))$ is generated by
\beq
C_4:=\bpm-1&0&\\0&-i&0\\0&0&1\epm
\eeq
in the same basis as before.  The refined Molien series is 
\beq\label{MSZ4}
P_{J_{\G_4}^{\suf(4)}}(t_1,t_2,t_3)
=\frac{1+t_1t_2^2}{(1-t_1^2)(1-t_2^4)(1-t_3)}.
\eeq
Its plethystic logarithm is
\beq
\cF^{\suf(4)}_{\G_4}(t_1,t_2,t_3)
=PE^{-1}\left(P_{J_{\G_4}^{\suf(4)}}\right)
=t_1^2+t_1t_2^2+t_2^4+t_3-t_1^2t_2^4 ,
\eeq
from which we can read off the coordinate ring
\beq
\C\{\cC_4(\suf(4))\}
=\C[\tu_1,\tu_2, \tu_3, \tu_4]/
\langle \tu_1\tu_2-\tu_4^2 \rangle
\eeq
where 
\begin{align}\label{}
\tu_1=u_1^2, \qquad
\tu_2=u_2^4, \qquad
\tu_3=u_3, \qquad\text{and}\qquad 
\tu_4=u_1u_2^2. 
\end{align}
So the generators of the daughter CB have scaling dimensions 4, 12, 4, and 8 respectively.  Again no CB parameters has scaling dimension 2 which is consistent with the theory having only $\cN=3$ supersymmetry.  The coordinate ring implies that the resulting CB, as a complex variety, is a hypersurface in $\C^4$:
\beq
\cC_4(\suf(4)):=
\Big\{
(\tu_1,\tu_2,\tu_3,\tu_4)\in\C^4
\big|
\tu_1\tu_2-\tu_4^2=0
\Big\}.
\eeq
As in the previous case the relations involve all of the coordinates but one, so again $\cC_4(\suf(4))$ has a one complex dimensional singular locus
\beq
\cV_{\cC_4(\suf(4))} =\Big\{(\tu_1,\tu_2,\tu_3,\tu_4)\in\C^4\Big|\tu_1=\tu_2=\tu_4=0\Big\}.
\eeq

\subsubsection{$\G_6$ gauging of the $[\SU(4)/\Z_2]_+$ theory}

To complete the analysis of the $[SU(4)/\Z_2]_+$ $\cN=4$ theory, let's compute the resulting CB after gauging the $\G_6$ symmetry,
\beq
\cC_6(\suf(4)) = \cC(\suf(4))/\G_6.
\eeq
The $\G_6$ action on $\cC(\suf(4))$ is generated by
\beq
\G_6:=\bpm e^{2\pi i/3}&0&\\
0&-1&0\\ 0&0& e^{4\pi i/3} \epm ,
\eeq
and the refined Molien series is then
\beq\label{MSZ3}
P_{J_{\G_6}^{\suf(4)}}(t_1,t_2,t_3)
=\frac{1+t_1t_3(1+t_1t_3)}{(1-t_1^3)(1-t_2^2)(1-t_3^3)},
\eeq
from which we compute the plethystic logarithm,
\beq
\cF^{\suf(4)}_{\G_6}(t_1,t_2,t_3)
=PE^{-1}\left(P_{J_{\G_6}^{\suf(4)}}\right)
=t_1^3+t_2^2+t_1t_3+t_3^3-t_1^3t_3^3,
\eeq
which in turn can be converted in an explicit expression for the coordinate ring of the resulting $\cN=3$ theory: 
\beq
\C\{\cC_6(\suf(4))\}=\C[\tu_1,\tu_2, \tu_3, \tu_4]/\langle \tu_1\tu_3-\tu_4^3\rangle ,
\eeq
where 
\begin{align}\label{}
\tu_1=u_1^3, \qquad
\tu_2=u_2^2, \qquad
\tu_3=u_3^3 \qquad\text{and}\qquad
\tu_4=u_1u_3. 
\end{align}
So the generators of the $\cN=3$ CB have scaling dimensions 6, 8, 12, and 6 respectively.  Again no CB parameter has scaling dimension 2, consistent with the theory having only $\cN=3$ supersymmetry.  The coordinate ring also implies that $\cC_6(\suf(4))$ is a hypersurface in $\C^4$,
\beq
\cC_6(\suf(4))
=
\Big\{(\tu_1,\tu_2,\tu_3,\tu_4)\in\C^4\big|
\tu_1\tu_3-\tu_4^3=0\Big\},
\eeq
with a one complex dimensional singular locus parametrized by $\tu_2$:
\beq
\cV_{\cC_6(\suf(4))} =\Big\{(\tu_1,\tu_2,\tu_3,\tu_4)\in\C^4\Big|\tu_1=\tu_3=\tu_4=0\Big\}.
\eeq

\subsection{$\cN=3$ theories with CB complex singularities and syzygies}

As we go up in rank, the complex structure of the CB of the daughter becomes quickly quite complicated.  The singular varieties thus far discussed could all be written as hypersurfaces in $\C^{r+1}$ but this is by no means the generic situation.  In fact it is easy to construct examples where the resultant geometry is not even a complete intersection.  Our final example will discuss a geometry of this type and involves a $\G_4$ gauging of the $\cN=4$ $\suf(5)$ sYM theory.

\subsubsection{$\G_4$ gauging of the $\suf(5)$ theory}

Consider the $\cN=4$ theory with gauge Lie algebra $\suf(5)$.  As mentioned earlier, all possible global forms of its gauge group and choices of its spectra of line operators are connected by S-dualities, so there is only one such theory.  As it was discussed at the end of section \ref{symms}, its S-duality group contains a $\Z_4$ factor and thus we can perform an $\cN{=}3$-preserving $\G_4$ gauging, giving the daughter CB
\beq\label{su5z4}
\cC_4(\suf(5)) = \cC(\suf(5))/\G_4.
\eeq
The action of a generator of $\G_4$ on $\cC(\suf(5))$ is given by
\beq
C_4 =\bpm
-1&0&0&0\\
0&-i&0&0\\
0&0&1&0\\
0&0&0&-1
\epm
\eeq
in a coordinate basis $(u_1,u_2,u_3,u_4)$ of $\cC(\suf(5))$ which have dimensions $(2,3,4,5)$, respectively.
Then the Molien series of the $\G_4$ action is given by
\beq\label{MSsu5Z4}
P_{J_{\G_4}^{\suf(5)}}(t_1,t_2,t_3,t_4)
=\frac{(1+t_2t_4)\big(1+t_1t_2^2+(t_1+t_2^2)t_4^2\big)}
{(1-t_1^2)(1-t_2^4)(1-t_3)(1-t_4^4)},
\eeq
which looks far from being in the factorized form \eqref{polci}.  Indeed, its plethystic logarithm gives
\begin{align}\label{gfsu5z4}
\cF^{\suf(5)}_{\G_4} 
= PE^{-1}\left(P_{J_{\G_4}^{\suf(5)}}\right) 
&=t_3+t_1^2+t_2t_4+t_1t_2^2+t_1t_4^2+t_2^4+t_4^4\\
& \qquad \mbox{}-t_1^2t_2^4-t_1^2t_2^2t_4^2-t_1^2t_4^4-t_1t_2^2t_4^4-t_1t_2^4t_4^2-t_2^4t_4^4+\cO(t^9)
\nonumber.
\end{align}
Before we write down the coordinate ring of the orbifold variety, let's discuss \eqref{gfsu5z4}.  In this case the generating function does not truncate, indicating a complex variety which cannot be written as a complete intersection.  A heuristic way to extract the generators and the relations from \eqref{gfsu5z4} is to order the series by the overall degree of each term as is done above.  We interpret the first consecutive positive signs as generators of the \eqref{su5z4} coordinate ring while the next terms coming with minus signs as relations among those generators.  We simply neglect the rest of the generating function.  Following this procedure we obtain a closed expression
\beq
\C\{\cC_4(\suf(5))\}
=\C[\tu_1,\ldots, \tu_7]/\cI_4(\suf(5))
\eeq
where 
\begin{align}\label{}
\tu_1=u_1^2, \ 
\tu_2=u_2^4, \ 
\tu_3=u_3, \ 
\tu_4=u_4^4, \ 
\tu_5=u_1u_2^2, \ 
\tu_6=u_1u_4^2 \  
\tu_7=u_2u_4,
\end{align}
and $\cI_4(\suf(5))$ is the ideal generated by six polynomials $\cU_i$ in the $\tu_i$'s:
\begin{align}\label{}
\cI_4(\suf(5)) 
= \vev{\cU_1,\ldots,\cU_6}
:=& \langle
\tu_1\tu_7^2 -\tu_5\tu_6,\ 
\tu_2\tu_6-\tu_5\tu_7^2,\ 
\tu_4\tu_5-\tu_6\tu_7^2,\nonumber\\
& \qquad\mbox{} \tu_5^2-\tu_1\tu_2,\
\tu_6^2-\tu_1\tu_4,\ 
\tu_7^4-\tu_2\tu_4
\rangle .
\end{align}
It is clear that $\cI_4(\suf(5))$ is not a free $\C[\tu_1,\ldots,\tu_7]$ module.  For instance $\tu_5\cU_1+\tu_1\cU_2+\tu_6\cU_4=0$; but in fact there are many relations.  We will make no attempt to study the syzygies of this coordinate ring and simply write down explicitly $\cC_4(\suf(5))$ as an algebraic variety embedded in affine $\C^7$:
\beq\label{su5compl}
\cC_4(\suf(5)):=\left\{(\tu_1,\ldots,\tu_7)\in\C^7
\ \Big| \ 
\cU_1=\cdots=\cU_6=0
\right\}.
\eeq
It is worth stressing that even though $\cC_4(\suf(5))$ is embedded in $\C^7$ via 6 algebraic relations, the resultant CB is still a rank 4 theory.  Relations among relations of the kind we pointed out above, show that the algebraic relations in \eqref{su5compl} are not all independent.  But we can't solve for any one relation in terms of the others either, so the presentation \eqref{su5compl} is the most economical one we can find.

Even in this case, none of the relations depend on $\tu_3$.  Thus $\cC_4(\suf(5))$ also has a one dimensional singular locus spanned by $\tu_3$,
\beq
\cV_{\cC_4(\suf(5))} 
=\Big\{(\tu_1,\ldots,\tu_7)\in\C^7
\ \Big|\ 
\tu_1=\tu_2=\tu_4=\tu_5=\tu_6=\tu_7=0
\Big\} .
\eeq

As this example clearly shows, CB geometries and their complex singularities can be made arbitrarily complicated.

\section{Open questions}\label{questions}

The Coulomb branch complex geometries of the new SCFTs constructed here show that the conjecture that all $\cN=2$ Coulomb branches have freely-generated holomorphic coordinate rings \cite{Tachikawa:2013kta,Beem:2014zpa} is false.  

It remains an open question of how generic CBs with complex singularities are within the class of all $\cN=2$ SCFTs.  In particular, could it be that the only examples of Coulomb branches with singular complex structures arise from gauging discrete symmetries of theories with regular Coulomb branches? If so maybe there is a refinement of the conjecture which could still characterize the complex structure of $\cN=2$ CBs? Or are there examples of consistent higher-rank Coulomb branch geometries whose complex singularities do not arise as orbifold singularities?

From the examples studied here it seems plausible that CBs of $\cN=2$ SCFTs can be arbitrarily complicated complex varieties.  We have in fact shown that, even just within the restricted set of discretely gauged theories, the CB of the daughter theory, as a complex algebraic variety, can be one of the following: isomorphic to $\C^r$; a hypersurface in $\C^{r+1}$; a complete intersection in $\C^{r+a}$; or an orbifold, non-complete intersection, algebraic variety.  In all cases but the first one, the CB has complex singularities.  

All singular CBs we constructed have non-isolated singularities.  It is unclear whether this is simply a common feature of the small sample of cases considered or if it is a generic feature either of discretely gauged theories or of $\cN=2$ CBs more broadly.
 
If complex singularities do represent a generic feature of CBs of $\cN=2$ SCFTs, why haven't we seen any example which is not a discretely gauged version of a theory with a regular CB?  An appealing possible explanation is that $\cN=2$ SCFTs with singular CBs form a distinct set under RG flows.  In \cite{Argyres:2017tmj} it was shown that the Riemann-Roch theorem implies that rank 1 theories with non-freely-generated CB coordinate rings necessarily flow to other such non-freely-generated theories under relevant deformations.  It has recently been shown in \cite{Caorsi:2018ahl} that all rank 1 CBs have freely-generated coordinate rings, so the rank 1 argument presented in \cite{Argyres:2017tmj} has no direct applicability.  But the kind of reasoning used there might generalize to higher rank and provide a nice explanation of why all the known methods --- through which many infinite families of examples of $\cN=2$ CBs have been constructed --- have failed to produce thus far an example with a non-freely-generated coordinate ring.

There are two obvious ways to extend the construction of new SCFTs discussed here.  The first is to stay with $\cN=4$ sYM parent theories, and construct symmetries along the lines outlined in section \ref{construct} but preserving only an $\cN=2$ supersymmetry.  Indeed, this has been discussed in the rank-1 case in some detail in \cite{am1604}, and is straight forward to generalize.

The second is to start instead with $\cN=2$ SCFTs which are gauge theories and to gauge their discrete $\cN{=}2$-preserving symmetries at weak coupling.  These discrete symmetries should be combinations of outer automorphisms of the gauge group together with some flavor automorphisms.  The experience in the rank-1 case \cite{am1604} suggests that there are constraints on what flavor automorphisms can be gauged (consistent with $\cN=2$ supersymmetry), but we do not understand precisely what those constraints are.\footnote{\emph{Note added:}  the recent paper \cite{Bourget:2018ond} discusses the theories resulting from gauging gauge group outer automorphisms of $\cN=2$ gauge theories.}  

A third, less obvious and more conjectural, way of extending the constructions of this paper does not rely on having a lagrangian description of the parent theory.  Instead, one may search for potential discrete symmetries of a strongly-coupled parent SCFT by looking for symmetries of the low energy effective action on its moduli space of vacua.  Although the symmetries identified in this way might just be accidental in the IR, evidence that they are exact may be gained by demanding consistency under relevant (e.g., mass) deformations.  This approach was pursued in the rank-1 case where the full set of possible RG flows could be probed, with positive results \cite{am1604}.  As mentioned in the last paragraph, it was often found that the consistent discrete symmetries found in this way involved flavor outer automorphisms in particular ways.  It would be interesting to see if this approach could be extended to higher-rank examples.

Another question is whether there are 't Hooft anomalies for some of the discrete symmetries discussed here which prevent gauging them while preserving $\cN=2$ supersymmetry.  Or, if not, do they have interesting implications for the symmetry group structure of the gauged theories as in \cite{Cordova:2018cvg,Seiberg:2018ntt}?

Finally, all the discrete symmetries discussed here act on the spectrum of BPS states in vacua out on the Coulomb branch through an action of the $\SL(2,\Z)$ S-duality group on their EM charge lattices of the low energy theory \cite{Kapustin:2006pk}.  The gauge-invariant operators creating these charged states are the Wilson and 't Hooft line operators in the low energy theory.  So should the S-duality symmetries discussed here also be 1-form symmetries acting on line operators?  If so, what effect does gauging them have on the spectrum of line operators of the resulting theory?

\acknowledgments

It is a pleasure to thank C. Beem, T. Bourton, J. Distler, M. Lemos, A. Pini, E. Pomoni, and L. Rastelli for useful discussions.  PCA was supported in part by DOE grant DE-SC0011784 and by Simons Foundation Fellowship 506770.  MM was supported in part by NSF grant PHY-1151392 and in part by NSF grant PHY-1620610.

\bibliographystyle{JHEP}
\providecommand{\href}[2]{#2}\begingroup\raggedright\endgroup

\end{document}